\documentclass[
    aps,prb,reprint
    ]{revtex4-2}
\usepackage{graphicx}
\usepackage{amsmath}
\usepackage{amssymb}
\usepackage{bm}
\usepackage{xcolor}
\usepackage{multirow}
\usepackage{hyperref}
\usepackage{physics2}
\usepackage[english]{babel}
\usephysicsmodule{ab, ab.braket}
\hypersetup{citecolor=blue, colorlinks = true, urlcolor = blue}

\begin{document}

\title{Symmetry-adapted models for multifold fermions with spin--orbit coupling}

\author{Koki Satow}
\affiliation{Department of Physics, Nagoya University, Nagoya 464-8602, Japan}

\author{Ai Yamakage}
\affiliation{Department of Physics, Nagoya University, Nagoya 464-8602, Japan}

\date{\today}

\begin{abstract}

    Multifold fermions, quasiparticles with multiple degeneracy protected by crystalline symmetries, exhibit a variety of intriguing phenomena stemming from their large topological charges and unique band structures.
    A comprehensive understanding of their response to external stimuli remains challenging, especially for types protected by nonsymmorphic symmetries where various degrees of freedom are intricately coupled. Here, we systematically construct effective models for multifold fermions that incorporate external fields based on crystalline symmetry. Specifically, we develop a $\bm{k} \cdot \bm{p}$ model for the threefold fermion protected by space group I2$_1$3 (No.~199) in the presence of spin--orbit coupling, and derive the terms for external fields. By complementing this with a tight-binding model, we investigate the magnetic field response and reveal the pair annihilation of magnetic monopoles. Furthermore, we construct a $\bm{k} \cdot \bm{p}$ model for the eightfold fermion in space group P$\bar{4}3n1'$ (No.~218), including its coupling to external fields. This work provides a robust theoretical foundation for advancing the study of external field responses and transport phenomena in multifold fermions, opening new avenues to explore their rich physics.

\end{abstract}

\maketitle

%%%%%%%%%%%%%%%%%%%%%%%%%%%%%%%%%%%%%%%%%%%%%%%%%%%%%%%%%%%%%%%%%%%%%%%%%%%%%%%%%%%%%%%%%
%%%%%%%%%%%%%%%%%%%%%%%%%%%%%%%%%%%%%%%% Section %%%%%%%%%%%%%%%%%%%%%%%%%%%%%%%%%%%%%%%%
%%%%%%%%%%%%%%%%%%%%%%%%%%%%%%%%%%%%%%%%%%%%%%%%%%%%%%%%%%%%%%%%%%%%%%%%%%%%%%%%%%%%%%%%%

\section{Introduction}

Recent progress in topological quantum materials has firmly established Weyl fermions as emergent spin-1/2 quasiparticles in solids \cite{yan2017-to,burkov2018-we,armitage2018-we,nagaosa2020-tr}.
Remarkably, crystalline symmetries open the door to quasiparticles with even higher pseudospin, enriching the landscape of attainable fermionic excitations in condensed-matter systems, which are absent in high-energy physics owing to Poincaré symmetry.
These are known as multifold fermions, exhibiting multiple band degeneracies—threefold, fourfold, sixfold, and eightfold—at high-symmetry points in the Brillouin zone, and can be distinguished by the irreducible representations of their space groups \cite{fang2012-mu,manes2012-ex,bradlyn2016-be,wieder2016-do,weng2016-to,tang2017-mu,chang2017-un,lu2019-ro,zhang2020-tw,inigo2024-mu}. Weyl fermions behave as sources and sinks of Berry curvature in momentum space, analogous to magnetic monopoles carrying a topological charge (Chern number) of $\pm 1$.
In contrast, multifold fermions possess larger topological charges of 2 or more compared to conventional Weyl fermions, and their unique band structures, arising from diverse degrees of freedom, lead to a variety of distinctive phenomena. These include double Fermi arcs on the surface \cite{tang2017-mu,chang2017-un,nandy2019-ge,sanchez2019-to,takane2019-ob,xiao2024-si}, the chiral anomaly \cite{ezawa2017-ch,cano2017-ch,balduini2024-in,mandal2025-ch,ahmad2025-lo}, unique optical responses \cite{sanchez2019-li,flicker2018-ch,kaushik2021-ma,dey2022-dy,ominato2025-th}, and characteristic transport phenomena \cite{tang2021-sp,kikuchi2022-qu,shen2023-ch,kikuchi2023-el,ma2024-an,nakazawa2025-no,kikuchi2025-ba}. Spin-1 fermions and double Weyl fermions, which have been experimentally identified \cite{huang2016-ne,takane2019-ob,rao2019-ob,soh2024-we}, are particularly well-studied, exhibiting transport phenomena originating from their distinctive band structures. For other types of multifold fermions, active material searches are ongoing in both theoretical and experimental research, and their existence has been theoretically predicted and, in some cases, experimentally confirmed in materials \cite{shen2023-ch,rong2023-re,meng2023-mu,inigo2024-mu,xiao2024-si}.

The theoretical study of multifold fermions has primarily focused on model construction through two main approaches: either by building lattice models for materials that have already been experimentally identified, such as CoSi, or by deriving $\bm{k} \cdot \bm{p}$ models and lattice models based on space group symmetries \cite{bradlyn2016-be,fulga2017-tr,nandy2019-ge,lu2019-ro,shi2020-sp}. Despite these efforts, research on multifold fermions has predominantly centered on certain types, like the experimentally confirmed threefold fermions, leaving the theoretical analysis of types for which candidate materials have not yet been identified largely underdeveloped. This gap is particularly pronounced for systems possessing nonsymmorphic symmetries, where the intricate coupling of diverse degrees of freedom complicates the theoretical treatment. For instance, while certain multifold fermions, such as the threefold fermion at the $\Gamma$ point in CoSi, can be described within a simpler framework involving symmorphic symmetries and neglecting spin, others demand the explicit inclusion of nonsymmorphic symmetries and spin degrees of freedom. Furthermore, constructing models that incorporate external fields, such as magnetic fields or strain, and accounting for the resulting changes in electronic states, has proven difficult. This is because conventional approaches often do not fully integrate space group symmetries, making it imperative to systematically construct Zeeman and strain terms that precisely adhere to the crystalline symmetries for accurate predictions.
Therefore, comprehensively addressing the diverse degrees of freedom in nonsymmorphic symmetry and constructing models that include external fields remains a significant but critical challenge for the further exploration of novel properties in multifold fermions.

In this paper, we systematically construct effective models for threefold and eightfold fermions with spin--orbit coupling (SOC) that incorporate external magnetic fields and strain by using the irreducible representation of the space group, in a similar way proposed in Ref.~\cite{kusunose2023-sa}.
We first derive the $\bm{k} \cdot \bm{p}$ Hamiltonian and a corresponding tight-binding model for the threefold fermion protected by space group I$2_13$ (No.~199) with SOC. Our model fully respects all crystalline symmetries, yielding a Zeeman term that differs from those in previously reported models \cite{bradlyn2016-be}. Combining these two complementary models, our analysis reveals how an applied magnetic field splits the threefold node into multiple magnetic monopoles and subsequently leads to their pair annihilation as the field strength increases. Furthermore, we extend our systematic approach to construct the $\bm{k} \cdot \bm{p}$ model for the eightfold fermion protected by space group P$\bar{4}3n1'$ (No.~218), including its response to external fields.

The paper is organized as follows. In Sec.~\ref{sec:3fold}, we focus on the threefold fermion protected by space group 199. We derive its low-energy effective $\bm k \cdot \bm p$ Hamiltonian and a corresponding tight-binding model, systematically including terms for external fields. We then investigate the response to an applied magnetic field, detailing the splitting of the threefold node into multiple magnetic monopoles and their annihilation. In Sec.~\ref{sec:8fold}, we apply our methodology to construct the $\bm k \cdot \bm p$ model for the eightfold fermion in space group 218. Section~\ref{sec:conclusion} provides a summary of our findings.
The appendices contain detailed derivations of the models and the group-theoretical analysis.

%%%%%%%%%%%%%%%%%%%%%%%%%%%%%%%%%%%%%%%%%%%%%%%%%%%%%%%%%%%%%%%%%%%%%%%%%%%%%%%%%%%%%%%%%
%%%%%%%%%%%%%%%%%%%%%%%%%%%%%%%%%%%%%%%% Section %%%%%%%%%%%%%%%%%%%%%%%%%%%%%%%%%%%%%%%%
%%%%%%%%%%%%%%%%%%%%%%%%%%%%%%%%%%%%%%%%%%%%%%%%%%%%%%%%%%%%%%%%%%%%%%%%%%%%%%%%%%%%%%%%%

\section{Threefold Fermion}
\label{sec:3fold}

In this section, we systematically construct an effective model for the threefold fermion based on its crystalline symmetries.
Threefold fermions with SOC emerge in space groups 199, 214, and 220 \cite{bradlyn2016-be}.
This paper focuses on the fermion in space group 199, which has the lowest symmetry among them and is a universal model.
The threefold fermion emerges at the P point, $\bm{k}_\text{P} = (\pi, \pi, \pi)$ in the conventional coordinate, protected by the symmetry of space group I$2_131'$ (No.~199). It is important to note that the P point is not a time-reversal-invariant momentum (TRIM).

%%%%%%%%%%%%%%%%%%%%%%%%%%%%%%%%%%%%%%% Subsection %%%%%%%%%%%%%%%%%%%%%%%%%%%%%%%%%%%%%%%

\subsection{Hamiltonian}

\subsubsection{\texorpdfstring{$k \cdot p$}{k.p} model}
\label{sec:kp_3fold_main}

%%%%%%%%%%%%%%%%%%%%%%%%%%%%%%%%%%%%%% Begin Table %%%%%%%%%%%%%%%%%%%%%%%%%%%%%%%%%%%%%%

\begin{table*}[t]
    \renewcommand{\arraystretch}{1.3}
    \caption{Basis of the irreps of momentum, magnetic field, strain, and $3 \times 3$ matrix expressions in the point group T. $\Gamma$ denotes irreps of the point group T. $u_{ij}$ are the strain tensors, $u_{ij} = \partial_j(\delta \bm R)_i $, where $\delta \bm R$ are local atomic displacements. The fifth column defines the notation for the $3 \times 3$ matrices corresponding to each irrep. The sixth column provides the explicit forms of $3 \times 3$ matrices for the symbols defined in the fifth column.}
    \begin{ruledtabular}
        \begin{tabular}{cccccc}
            $\Gamma$ & Momentum        & Magnetic field  & Strain                                          & Symbol                               & Definition                                                                                                                                                                                      \\
            \hline
            $A$      & -               & -               & $u_{xx}+u_{yy}+u_{zz}$                          & $S^A$                                & \footnotesize $\begin{pmatrix} 1&0&0\\0&1&0\\0&0&1 \end{pmatrix}$                                                                                                                               \\
            $E$      & -               & -               & $[u_{xx}-u_{yy},2u_{zz}-u_{xx}-u_{yy}]$         & $[S_{x^2-y^2}^E,S_{2z^2-x^2-y^2}^E]$ & \footnotesize $\left[ \begin{pmatrix} 1&0&0\\0&-2&0\\0&0&1 \end{pmatrix}, \, \begin{pmatrix} 1&0&0\\0&0&0\\0&0&-1 \end{pmatrix} \right]$                                                        \\
            $T$      & $[k_x,k_y,k_z]$ & $[B_x,B_y,B_z]$ & $[u_{xy}+u_{yx}, u_{zx}+u_{xz}, u_{yz}+u_{zy}]$ & $[S_{1x}^T,S_{1y}^T,S_{1z}^T]$       & \footnotesize $\left[ \begin{pmatrix} 0&0&0\\0&0&1\\0&1&0  \end{pmatrix}, \, \begin{pmatrix} 0&0&-1\\0&0&0\\-1&0&0 \end{pmatrix}, \, \begin{pmatrix} 0&1&0\\1&0&0\\0&0&0 \end{pmatrix} \right]$ \\
                     &                 &                 & $[u_{xy}-u_{yx}, u_{zx}-u_{xz}, u_{yz}-u_{zy}]$ & $[S_{2x}^T,S_{2y}^T,S_{2z}^T]$       & \footnotesize $\left[ \begin{pmatrix} 0&0&0\\0&0&-i\\0&i&0 \end{pmatrix}, \, \begin{pmatrix} 0&0&-i\\0&0&0\\i&0&0 \end{pmatrix}, \, \begin{pmatrix} 0&-i&0\\i&0&0\\0&0&0 \end{pmatrix} \right]$
        \end{tabular}
    \end{ruledtabular}
    \label{tab:T_kBS}
\end{table*}

%%%%%%%%%%%%%%%%%%%%%%%%%%%%%%%%%%%%%%%%%%%%%%%%%%%%%%%%%%%%%%%%%%%%%%%%%%%%%%%%%%%%%%%%%

This section details the construction of the $\bm{k} \cdot \bm{p}$ Hamiltonian for the threefold fermion, considering the application of an external magnetic field. The $\bm{k} \cdot \bm{p}$ Hamiltonian for the threefold fermion can be expressed as $H(\bm{k}) = \sum_{i=1}^9 f_i(\bm{k}) S_i$, where $\bm{k}$ denotes the momentum measured from the P point, and $f_i(\bm{k})$ are polynomials in $\bm{k}$ (up to linear order in this work), while $S_i$ are $3 \times 3$ Hermitian matrices. This Hamiltonian must satisfy the symmetry relation $gH(\bm{k})g^{-1} = H(g\bm{k})$ under operations $g$ of the little group $G^{\bm{k}_\mathrm{P}}_{199}$ at the P point.
By classifying the polynomials $f_i(\bm{k})$ and the matrices $S_i$ according to the bases of the irreducible representations of point group T, which comprises the rotational part of $G_{\bm{k}_\mathrm{P}}^{199}$, the Hamiltonian can be systematically constructed. Detailed calculations are provided in Appendix~\ref{sec:kp_3fold}. Consequently, the effective $\bm{k} \cdot \bm{p}$ Hamiltonian for the threefold fermion, $H_{199,\mathrm{P}} = \sum_{\bm{k}} c_{\bm{k}}^\dagger H_{199,\mathrm{P}}(\bm{k}) c_{\bm{k}}$, is given by
\begin{align}
    H_{199,\mathrm P}(\bm k) & = \bm k \cdot (v_1 \bm S_1^{T} + v_2 \bm S_2^{T}).
\end{align}
where $v_1$ and $v_2$ are arbitrary real coefficients, and $\bm S_1^T = ( S_{1x}^T, S_{1y}^T, S_{1z}^T )$ and $\bm S_2^T= ( S_{2x}^T, S_{2y}^T, S_{2z}^T )$ serve as basis matrices, as defined in Table~\ref{tab:T_kBS}. We can verify that $\boldsymbol{k}$, $\boldsymbol{S}_1^T$, and $\boldsymbol{S}_2^T$ belong to the same irreducible representation $T$; hence, the above Hamiltonian is totally symmetric. The Hamiltonian can similarly be extended to account for external fields. Table~\ref{tab:T_kBS} summarizes the irreducible decomposition for both magnetic fields and strains.  Based on this, the Zeeman term for an applied magnetic field $\bm{B}$ is derived as
\begin{align}
    H_{199,\mathrm P}^{\mathrm Z}(\bm B) & = \mu_{\mathrm B} \bm B \cdot (g_1 \bm S_1^{T} + g_2 \bm S_2^{T}).
\end{align}
This Zeeman term rigorously incorporates all crystalline symmetries of space group $\mathrm I2_131'$ (No.~199) that are currently under consideration, unlike that presented in the previous model \cite{bradlyn2016-be}.

Furthermore, when considering the strain term to the lowest order, it is explicitly given by $H_{199,\mathrm P}^{\mathrm{strain}}(u) = \int d^3 x \psi^\dagger(\bm x) (h^A + h^E + h^T) \psi(\bm x)$. Here, $h^{\Gamma}$ represents a totally symmetric representation constructed from the product representation $\Gamma \times \Gamma$ of the strain tensor $u_{ij} = \partial_j(\delta \bm R)_i$, where $\delta \boldsymbol{R}$ denotes the local displacement field, and the $3 \times 3$ matrices $S_i$'s. They belong to the same irreducible representation $\Gamma$ of the point group T. As a specific example, for the case of $\Gamma = E$, the term $h^E$ is explicitly written as
\begin{align}
    h^E = & (u_{xx} - u_{yy}) S_{x^2-y^2}^E  \notag               \\
          & + (2u_{zz} - u_{xx} - u_{yy}) S_{2z^2 - x^2 - y^2}^E.
\end{align}

\subsubsection{Tight-binding model}

%%%%%%%%%%%%%%%%%%%%%%%%%%%%%%%%%%%%%% Begin Figure %%%%%%%%%%%%%%%%%%%%%%%%%%%%%%%%%%%%%%

\begin{figure*}[t]
    \includegraphics[width=0.8\textwidth,clip]{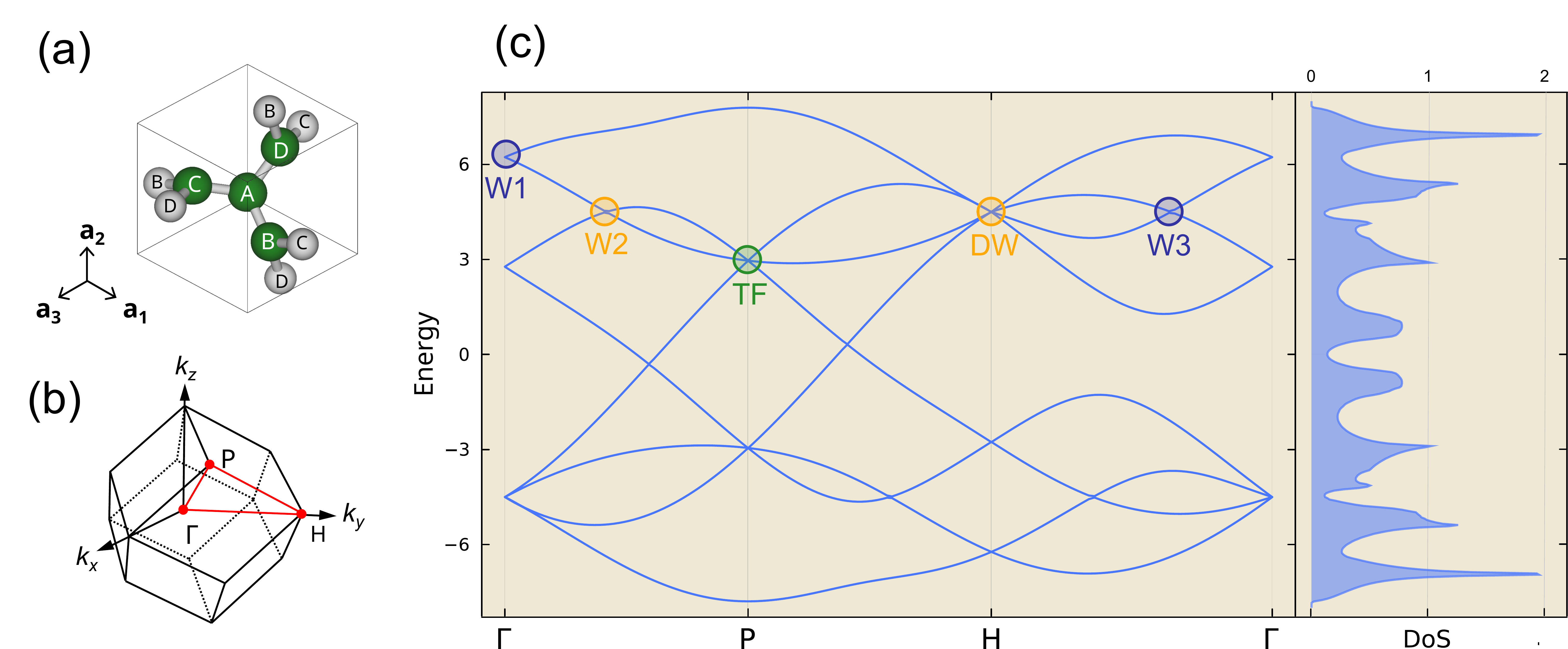}%
    \caption{(a) Primitive unit cell which respects $I2_13$ symmetry \cite{Momma2011-ve}. The green sites are included in the cell, while the gray sites are located in the next cells. The primitive lattice vectors of bcc, $\bm{a}_1, \bm{a}_2$, and $\bm{a}_3$, are indicated.
        (b) Brillouin zone. (c) Band structure and density of states (DOS). The parameters are taken as $\lambda_1 = 2.0, \, \lambda_2 = 1.0, \, \lambda_3 = -1.5$.}
    \label{fig:tb199}
\end{figure*}

%%%%%%%%%%%%%%%%%%%%%%%%%%%%%%%%%%%%%%%%%%%%%%%%%%%%%%%%%%%%%%%%%%%%%%%%%%%%%%%%%%%%%%%%%

%%%%%%%%%%%%%%%%%%%%%%%%%%%%%%%%%%%%%% Begin Table %%%%%%%%%%%%%%%%%%%%%%%%%%%%%%%%%%%%%%

\begin{table*}[t]
    \renewcommand{\arraystretch}{1.6}
    \caption{Basis of the irreps of spin, magnetic field, on-site interactions, and NN hopping in the point group T. $\Gamma$ denotes irreps of the point group T. The fifth column defines the notation for the on-site energies and NN hopping, $\bm \epsilon = (u_{\mathrm{AA}},u_{\mathrm{BB}},u_{\mathrm{CC}},u_{\mathrm{DD}}), \, \bm t^{\Gamma} = ( t_{\mathrm{AB}}^\Gamma, t_{\mathrm{AC}}^\Gamma, t_{\mathrm{AD}}^\Gamma, t_{\mathrm{BC}}^\Gamma, t_{\mathrm{BD}}^\Gamma, t_{\mathrm{CD}}^\Gamma )$, corresponding to each irrep. The sixth column provides the explicit on-site interactions and NN hopping for the symbols defined in the fifth column.
    }
    \begin{ruledtabular}
        \begin{tabular}{cccccc}
            $\Gamma$ & TR  & Spin                           & Magnetic field  & Symbol                                                             & Definition                                                                                                                                                                          \\
            \hline
            $A$      & +   & $\sigma_0$                     & -               & $\bm t^{A^+}$                                                      & $(1,1,1,1,1,1)$                                                                                                                                                                     \\
                     &     &                                &                 & $\bm \epsilon^{A^+}$                                               & $(1,1,1,1)$                                                                                                                                                                         \\
            $E$      & $-$ & -                              & -               & $[\bm t_{x^2-y^2}^{E^-},\bm t_{2z^2-x^2-y^2}^{E^-}]$               & $[(1,e^{i\frac{2\pi}{3}},e^{-i\frac{2\pi}{3}},e^{-i\frac{2\pi}{3}},e^{i\frac{2\pi}{3}},1),(1,e^{-i\frac{2\pi}{3}},e^{i\frac{2\pi}{3}},e^{i\frac{2\pi}{3}},e^{-i\frac{2\pi}{3}},1)]$ \\
            $T$      & +   & -                              & -               & $[\bm t_x^{T^+}, \bm t_y^{T^+}, \bm t_z^{T^+}]$                    & $[(0,0,1,-1,0,0), \, (0,1,0,0,-1,0), \, (1,0,0,0,0,-1)]$                                                                                                                            \\
                     &     &                                &                 & $[\bm \epsilon_x^{T^+},\bm \epsilon_y^{T^+},\bm \epsilon_z^{T^+}]$ & $[(1,1,-1,-1),(1,-1,-1,1),(1,-1,1,-1)]$                                                                                                                                             \\
                     & $-$ & $[\sigma_x,\sigma_y,\sigma_z]$ & $[B_x,B_y,B_z]$ & $[\bm t_{1x}^{T^-}, \bm t_{1y}^{T^-}, \bm t_{1z}^{T^-}]$           & $[(i,0,0,0,0,-i), \, (0,0,i,-i,0,0), \, (0,i,0,0,i,0)] $                                                                                                                            \\
                     &     &                                &                 & $[\bm t_{2x}^{T^-}, \bm t_{2y}^{T^-}, \bm t_{2z}^{T^-}]$           & $[(0,i,0,0,-i,0), \, (i,0,0,0,0,i), \, (0,0,i,i,0,0)]$                                                                                                                              \\
        \end{tabular}
    \end{ruledtabular}
    \label{tab:T_ts}
\end{table*}

%%%%%%%%%%%%%%%%%%%%%%%%%%%%%%%%%%%%%%%%%%%%%%%%%%%%%%%%%%%%%%%%%%%%%%%%%%%%%%%%%%%%%%%%%

This section describes the construction of a tight-binding model for the threefold fermion, which respects the symmetry of the space group.
First, we construct a system that possesses space group $\mathrm I2_131'$ (No.~199) symmetry by placing an $s$-orbital with spin 1/2 at Wyckoff position 8a \footnote{Note that any orbital on the 8a site induces threefold degeneracy at the P point, according to BANDREP \cite{Aroyo2011-cr,Aroyo2006-bi1,Aroyo2006-bi2,bradlyn2017-to,vergniory2017-gr,elcoro2017-do}}.
The primitive unit cell contains four sublattices, denoted by A, B, C, and D. The lattice structure is illustrated in Fig.~\ref{fig:tb199}(a), and its Brillouin zone is shown in Fig.~\ref{fig:tb199}(b). The Hamiltonian must be totally symmetric and is therefore given as
\begin{align}
     &
    H = \sum_{\boldsymbol{k}} \bm c_{\boldsymbol{k}}^\dag H_{\boldsymbol{k}} \bm c_{\boldsymbol{k}},
    \\
     &
    H_{\boldsymbol{k}} = \sum_{\Gamma} \boldsymbol{t}_{\boldsymbol{k}}^\Gamma \cdot \boldsymbol{\sigma}^\Gamma,
    \\
     &
    \ab(\boldsymbol{t}_{\boldsymbol{k}}^\Gamma)_{XY} = \ab(\boldsymbol{t}^\Gamma)_{XY} e^{i \boldsymbol{k} \cdot \boldsymbol{d}_{XY}}.
\end{align}
We define the row vector $c_{\boldsymbol{k}}^\dagger$ by arranging the fermion operators $c_{\boldsymbol{k} X, i}^\dagger$, where $c_{\boldsymbol{k} X, i}^\dagger$ creates an electron with spin $i$ at sublattice $X$. Here, $\Gamma$ denotes the irreducible representation of point group T, which is compatible with space group 199. The vectors $\bm t^\Gamma = (\bm t_1^\Gamma, \cdots, \bm t_{d}^\Gamma)^\top$ and $\bm \sigma^\Gamma = (\bm \sigma_1^\Gamma, \cdots, \bm\sigma_d^\Gamma)^\top$ correspond to the basis of the $d$-dimensional irreducible representation $\Gamma$ for the hopping and the spin, respectively. The term $t_{{XX}}$ represents the onsite energy of sublattice $X$, while $t_{{XY}}$ represents the nearest-neighbor (NN) hopping between sublattices $X$ and $Y$. The matrices $\sigma_0$, $\sigma_x$, $\sigma_y$, and $\sigma_z$ are the identity and Pauli matrices for the spin degrees of freedom, respectively. The vector $\bm d_{{XY}}$ points from sublattice $X$ to its nearest-neighbor $Y$ sublattice. The hopping terms $\bm{t}$ and spin matrices $\bm\sigma$ are decomposed into the basis of irreducible representation under the point group T. The results are presented in Table~\ref{tab:T_ts}, with detailed calculations provided in Appendix~\ref{sec:3fold_tb}. Constructing a totally symmetric representation from the product representation of the $\bm{t}$ and $\bm{\sigma}$, the Hamiltonian invariant under space group $\mathrm I2_131'$ (No.~199) symmetry is given as
\begin{align}
    H_{199 \bm k} = \lambda_1 t_{\bm k}^{A^+} + \lambda_2 \bm t_{1 \bm k}^{T^-} \cdot \bm \sigma + \lambda_3 \bm t_{2 \bm k}^{T^-} \cdot \bm \sigma.
    \label{H199k}
\end{align}
Additionally, the Zeeman term, when a magnetic field $\boldsymbol{B}$ is applied, is given by
\begin{align}
    H_{199 \boldsymbol{k}}^{\mathrm Z}(\bm B) & = \mu_{\mathrm{B}} \bm B \cdot \bm \sigma.
\end{align}

%%%%%%%%%%%%%%%%%%%%%%%%%%%%%%%%%%%%%%% Subsection %%%%%%%%%%%%%%%%%%%%%%%%%%%%%%%%%%%%%%%

\subsection{Splitting of Monopoles by Applying a Magnetic Field}

Weyl points can be controlled by tuning external parameters, which can lead to their pair creation or annihilation \cite{cano2017-ch,zhang2017-ma,ghimire2019-cr,konye2021-ch,naselli2024-st}. In the context of multifold fermions, applying external fields, such as magnetic fields or strain, breaks the symmetries that protect the threefold fermion, causing it to split into multiple magnetic monopoles. In this section, we investigate the splitting of the threefold fermion upon applying a magnetic field as a Zeeman term, utilizing both the $\bm k \cdot \bm p$ model and the tight-binding model.
The parameters $v_1$, $v_2$, $g_1$, and $g_2$ of the $\bm k \cdot \bm p$ model, which describes the threefold fermion appearing at the P point, are related to the tight-binding model parameters $\lambda_1$, $\lambda_2$, and $\lambda_3$ by the following expressions
\begin{align}
    v_1 & = \frac{1}{4} ( -\lambda_1 - \lambda_2 + \lambda_3), \label{eq:v1}                                                                                                                                                                                                       \\
    v_2 & = \pm \frac{ \lambda_1^2 + \lambda_2^2 + \lambda_3^2 + 2(- \lambda_1 \lambda_2 + \lambda_1 \lambda_3 + \lambda_2 \lambda_3)}{4 \sqrt{3( \lambda_1^2 + \lambda_2^2 + \lambda_3^2 ) + 2(-\lambda_1 \lambda_2 + \lambda_1 \lambda_3 + \lambda_2 \lambda_3)}}, \label{eq:v2} \\
    g_1 & = \pm \frac{ -3\lambda_1 + \lambda_2 - \lambda_3 }{2 \sqrt{3( \lambda_1^2 + \lambda_2^2 + \lambda_3^2 ) + 2(-\lambda_1 \lambda_2 + \lambda_1 \lambda_3 + \lambda_2 \lambda_3)}}, \label{eq:g1}                                                                           \\
    g_2 & = \frac{1}{2}. \label{eq:g2}
\end{align}
Here, the positive and negative signs correspond to the higher- and lower-energy threefold fermions at the P point in the tight-binding model, respectively.
The derivation of these relations is detailed in Appendix~\ref{sec:3fold_correspondence}.
For the calculations presented in this section, we set $\lambda_1=2.0$, $\lambda_2=1.0$, and $\lambda_3=-1.5$.

\begin{figure}[htbp]
    \includegraphics[scale=0.38]{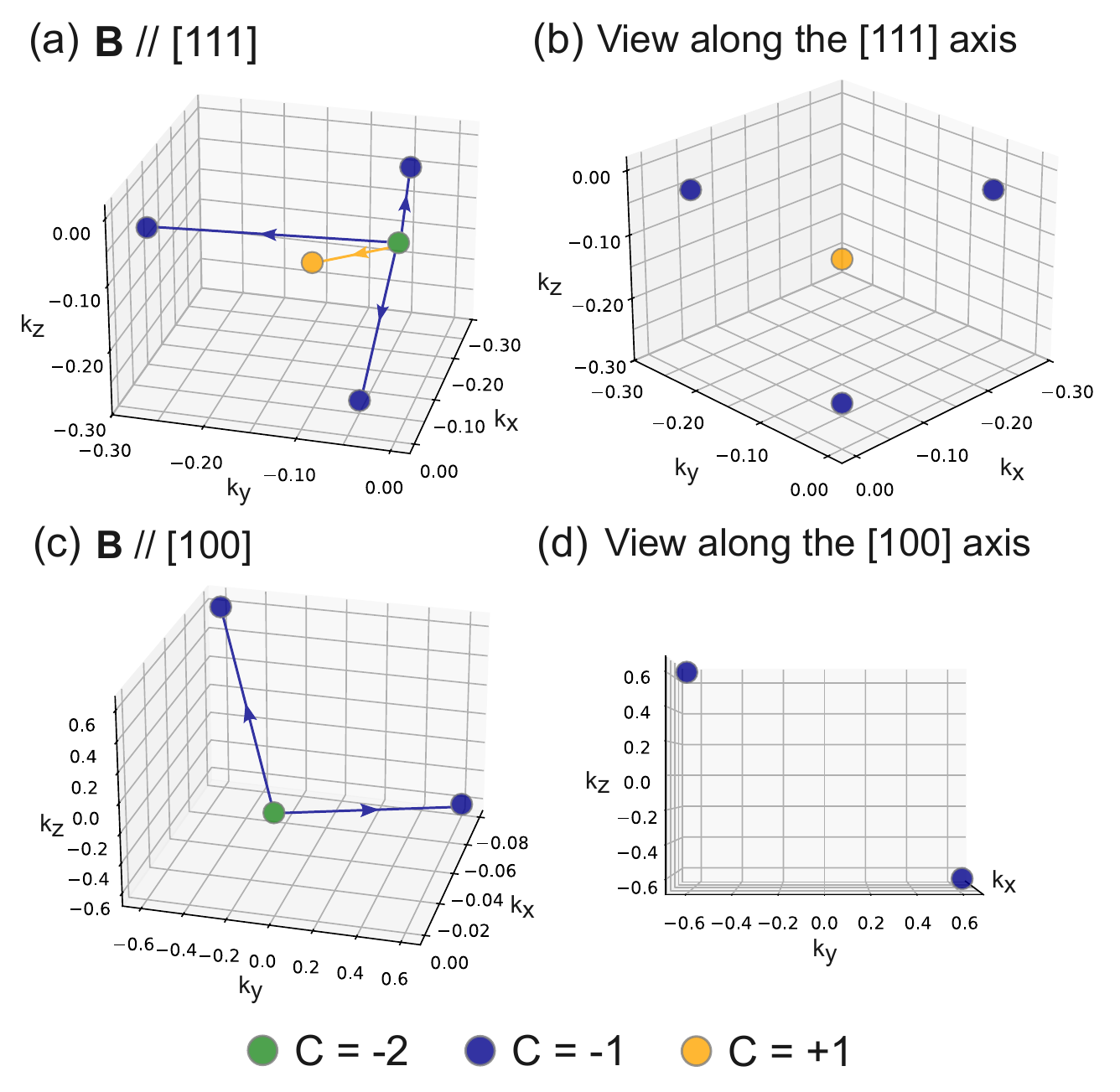}
    \caption{Splitting of magnetic monopoles in the $\bm k \cdot \bm p$ model when a magnetic field is applied along [(a) and (b)] the [111] axis and [(c) and (d)] the [100] axis.
        (a) and (c) are shown for both cases with and without the field, while (b) and (d) are shown for the case with the field.
        The parameters $v_1$, $v_2$, $g_1$, and $g_2$ are obtained from Eqs.~\eqref{eq:v1}--\eqref{eq:g2} using $\lambda_1=2.0$, $\lambda_2=1.0$, and $\lambda_3=-1.5$.}
    \label{fig:kpmonopoles}
\end{figure}

Figure~\ref{fig:kpmonopoles} illustrates the splitting behavior of the threefold fermion within the $\bm k \cdot \bm p$ model when a magnetic field is applied along high-symmetry lines.
The monopole charge for each degenerate point is calculated for the highest band by utilizing the Fukui--Hatsugai--Suzuki method \cite{fukui2005-ch}.
When the magnetic field is applied along the [111] axis, the threefold degeneracy with monopole charge $C=-2$ splits into four magnetic monopoles: a triplet of nodes, each with a charge of $C=-1$, and one monopole with a charge of $C=+1$.
Consistent with the Nielsen--Ninomiya theorem, the total charge is conserved to be $-2$.
When the magnetic field is applied along the [100] axis, the threefold fermion splits into two magnetic monopoles, each carrying a charge of $C=-1$.
The configurations of magnetic monopoles respect the remaining crystalline symmetries under the magnetic field, satisfying $C_{3,111}$ symmetry for $\boldsymbol{B} \parallel [111]$ in Fig.~\ref{fig:kpmonopoles}(b) and $C_{2x}$ symmetry for $\boldsymbol{B} \parallel [100]$ in Fig.~\ref{fig:kpmonopoles}(d).

%%%%%%%%%%%%%%%%%%%%%%%%%%%%%%%%%%%%%% Begin Figure %%%%%%%%%%%%%%%%%%%%%%%%%%%%%%%%%%%%%%

%%%%%%%%%%%%%%%%%%%%%%%%%%%%%%%%%%%%%%%%%%%%%%%%%%%%%%%%%%%%%%%%%%%%%%%%%%%%%%%%%%%%%%%%%

\begin{figure}[htbp]
    \includegraphics[width=0.48\textwidth,clip]{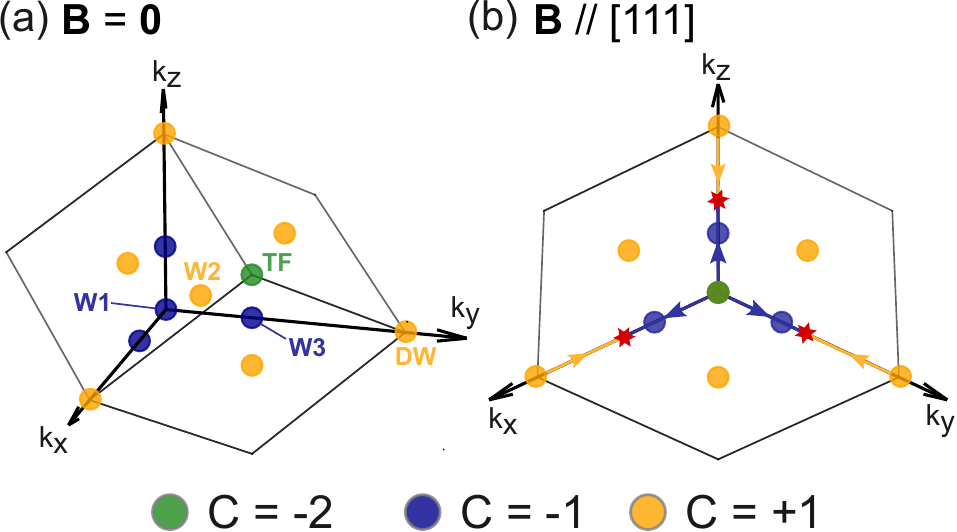}
    \caption{Splitting of magnetic monopoles by a magnetic field in the Brillouin zone for a lattice system with space group $\mathrm{I}2_131'$ (No.~199) symmetry.
        The monopole charge is calculated for the second band from the highest in energy, which is the highest among the threefold fermions.
        (a) The magnetic monopoles without a magnetic field. The labels, TF, DW, W1, W2, and W3, correspond to those in Fig.~\ref{fig:tb199}.
        (b) Splitting of the threefold fermion at the P point when a magnetic field is applied along the [111] axis.
        As the magnetic field increases, the magnetic monopoles shift as indicated by the arrows, then annihilate in pairs at the points denoted by the red stars.
        A $C=+1$ magnetic monopole, which splits from the threefold fermion, is on the $\Gamma$-P line (not shown).}
    \label{fig:tbmonopoles}
\end{figure}

Next, we investigate the splitting behavior of magnetic monopoles across the entire Brillouin zone using the tight-binding model. Figure~\ref{fig:tbmonopoles}(a) plots the configuration of magnetic monopoles in the Brillouin zone within the region where $k_x, k_y, k_z>0$ when $\bm B = \bm 0$.
For this analysis, the monopole charge is calculated for the second band from the highest in energy, which is the highest among the threefold fermions.
Upon applying a magnetic field, the $C=-2$ threefold fermion located at the P point splits into four magnetic monopoles, consistent with the behavior observed in the $\bm k \cdot \bm p$ model. Furthermore, the tight-binding model allows us to investigate the pair annihilation of magnetic monopoles as the magnetic field strength increases, a phenomenon that could not be captured by the $\bm k \cdot \bm p$ model alone. For example, when a magnetic field is applied along the [111] axis, the $C=-1$ magnetic monopoles, which form a triplet upon splitting from the threefold fermion, each annihilate with a $C=+1$ magnetic monopole initially located at the H point [see Fig.~\ref{fig:tbmonopoles}(b)].

%%%%%%%%%%%%%%%%%%%%%%%%%%%%%%%%%%%%%% Begin Figure %%%%%%%%%%%%%%%%%%%%%%%%%%%%%%%%%%%%%%

%%%%%%%%%%%%%%%%%%%%%%%%%%%%%%%%%%%%%%%%%%%%%%%%%%%%%%%%%%%%%%%%%%%%%%%%%%%%%%%%%%%%%%%%%

%%%%%%%%%%%%%%%%%%%%%%%%%%%%%%%%%%%%%%%%%%%%%%%%%%%%%%%%%%%%%%%%%%%%%%%%%%%%%%%%%%%%%%%%%
%%%%%%%%%%%%%%%%%%%%%%%%%%%%%%%%%%%%%%%% Section %%%%%%%%%%%%%%%%%%%%%%%%%%%%%%%%%%%%%%%%
%%%%%%%%%%%%%%%%%%%%%%%%%%%%%%%%%%%%%%%%%%%%%%%%%%%%%%%%%%%%%%%%%%%%%%%%%%%%%%%%%%%%%%%%%

\section{Eightfold Fermion}
\label{sec:8fold}

%%%%%%%%%%%%%%%%%%%%%%%%%%%%%%%%%%%%%% Begin Table %%%%%%%%%%%%%%%%%%%%%%%%%%%%%%%%%%%%%%

\begin{table}[t]
    \caption{Basis of the irreps of momentum, magnetic field, and strain in the point group $\mathrm T_d$.}
    \begin{ruledtabular}
        \begin{tabular}{ccc}
            $\Gamma$ & $\bm k, \, \bm B$ & Strain                                          \\
            \hline
            $A_1$    & -                 & $u_{xx}+u_{yy}+u_{zz}$                          \\
            $A_2$    & -                 & -                                               \\
            $E$      & -                 & $[2u_{zz}-u_{xx}-u_{yy},u_{xx}-u_{yy}]$         \\
            $T_1$    & $[B_x,B_y,B_z]$   & -                                               \\
            $T_2$    & $[k_x,k_y,k_z]$   & $[u_{xy}+u_{yx}, u_{zx}+u_{xz}, u_{yz}+u_{zy}]$ \\
                     &                   & $[u_{xy}-u_{yx}, u_{zx}-u_{xz}, u_{yz}-u_{zy}]$
        \end{tabular}
    \end{ruledtabular}
    \label{tab:Td_kb}
\end{table}

%%%%%%%%%%%%%%%%%%%%%%%%%%%%%%%%%%%%%%%%%%%%%%%%%%%%%%%%%%%%%%%%%%%%%%%%%%%%%%%%%%%%%%%%%

Eightfold fermions with SOC emerge in space group 130, 135, 218, 220, 222, 223, and 230 \cite{bradlyn2016-be}. This paper focuses on the fermion in space group 218, as it possesses the simplest symmetry among those hosted in a cubic primitive Bravais lattice. The eightfold fermion emerges at the R point, $\bm{k}_\text{R} = (\pi, \pi, \pi)$ in the conventional coordinate, protected by the symmetry of space group P$\bar{4}3n1'$ (No.~218). In this section, we systematically construct the $\bm k \cdot \bm p$ Hamiltonian for the eightfold fermion based on its crystalline symmetries. The character table for $\mathrm T_d$, which comprises the rotational part of the little group at the R point, along with the basis of momentum $\bm k$ and magnetic field $\bm B$, is presented in Table~\ref{tab:Td_kb}.
The results of the irreducible decomposition of $8 \times 8$ matrices are presented in Appendix~\ref{sec:Td_basis}, with their detailed derivations provided in Appendix~\ref{sec:kp_8fold}.
The $\bm k \cdot \bm p$ Hamiltonian for the eightfold fermion is given by
\begin{align}
    H_{218}(\bm k)
    = \sum_{\alpha=1}^6 v_\alpha \boldsymbol{k} \cdot \boldsymbol{S}_{\alpha}^{T_2^-},
\end{align}
where $v_\alpha$ is a real coefficient. The Zeeman term is expressed as
\begin{align}
    H_{218}^{\mathrm{Z}}(\bm B)
    = \mu_B \sum_{\alpha=1}^4 g_\alpha \boldsymbol{B} \cdot \boldsymbol{S}_{\alpha}^{T_1^-},
\end{align}
where $g_\alpha$ is a real coefficient.

Furthermore, the strain term to the lowest order is given by $H_{218,\mathrm R}^{\mathrm{strain}}(u) = \int d^3 x \psi^\dagger(\bm x) (h^A + h^E + h^{T_2}) \psi(\bm x)$, following the same procedure as in Sec.~\ref{sec:kp_3fold_main}. As a specific example, for the case of $\Gamma = E$, the term $h^E$ is explicitly written as
\begin{align}
    h^E = & (2u_{zz} - u_{xx} - u_{yy}) S_{2z^2 - x^2 - y^2}^{E^+} \notag \\
          & + (u_{xx} - u_{yy}) S_{x^2-y^2}^{E^+}.
\end{align}

%%%%%%%%%%%%%%%%%%%%%%%%%%%%%%%%%%%%%%%%%%%%%%%%%%%%%%%%%%%%%%%%%%%%%%%%%%%%%%%%%%%%%%%%%
%%%%%%%%%%%%%%%%%%%%%%%%%%%%%%%%%%%%%%%% Section %%%%%%%%%%%%%%%%%%%%%%%%%%%%%%%%%%%%%%%%
%%%%%%%%%%%%%%%%%%%%%%%%%%%%%%%%%%%%%%%%%%%%%%%%%%%%%%%%%%%%%%%%%%%%%%%%%%%%%%%%%%%%%%%%%

\section{Conclusions}
\label{sec:conclusion}

In this work, we have developed symmetry-adapted models for multifold fermions, with a focus on threefold and eightfold degeneracies, including an external field. The $\bm k \cdot \bm p$ model, represented by an $n \times n$ matrix for an $n$fold fermion, is the minimal effective theory that reveals the universal properties intrinsic to multifold quasiparticles.
It shows that a threefold fermion with monopole charge $-2$ splits into four nodes under a magnetic field along the $[111]$ direction, whereas that splits into two nodes when the field is along $[100]$--a behavior irrespective of material details.
To capture responses that depend on the entire Brillouin zone, however, a tight‑binding model is more suitable.
Unlike the $\bm k \cdot \bm p$ approach, the tight‑binding framework enforces a net magnetic charge of zero and can therefore describe how magnetic monopoles ultimately annihilate as the field strength increases.
Used together, these two complementary models provide a deeper, more essential understanding of phenomena arising from multifold fermions.
Our findings contribute to the ongoing exploration of unconventional quasiparticles in condensed matter systems and offer a theoretical foundation for predicting and interpreting experimental observations in topological quantum materials hosting such high-fold degeneracies.

\acknowledgments

This work is supported by JSPS KAKENHI for Grants (Grants Nos.~JP24H00853 and JP25K07224).

\appendix

%%%%%%%%%%%%%%%%%%%%%%%%%%%%%%%%%%%%%%%%%%%%%%%%%%%%%%%%%%%%%%%%%%%%%%%%%%%%%%%%%%%%%%%%%
%%%%%%%%%%%%%%%%%%%%%%%%%%%%%%%%%%%%%%%% Section %%%%%%%%%%%%%%%%%%%%%%%%%%%%%%%%%%%%%%%%
%%%%%%%%%%%%%%%%%%%%%%%%%%%%%%%%%%%%%%%%%%%%%%%%%%%%%%%%%%%%%%%%%%%%%%%%%%%%%%%%%%%%%%%%%

\section{\texorpdfstring{$\bm k \cdot \bm p$}{k.p} model for threefold fermions under magnetic field}

%%%%%%%%%%%%%%%%%%%%%%%%%%%%%%%%%%%%%%% Subsection %%%%%%%%%%%%%%%%%%%%%%%%%%%%%%%%%%%%%%%

\subsection{Threefold fermion at P point}
\label{sec:kp_3fold}

In this section, we construct the $\bm k \cdot \bm p$ model for the threefold fermion under an external magnetic field at the P point. The Hamiltonian for the threefold fermion is expressed as $H(\bm k) = \sum_{i=1}^{9} f_i(\bm k)S_i$, where $f_i(\bm k)$ are polynomials in momentum $\bm k$, and $S_i$ are linearly independent $3 \times 3$ Hermitian matrices. In this paper, we consider $f_i(\bm k)$ to be linear polynomials in momentum.
At the P point, $\bm k_{\mathrm P}=(\pi,\pi,\pi)$ in SG 199 ($G_{199}$), the little group $G^{\bm k_{\mathrm P}}_{199} \subset G_{199}$ hosts a double-valued three-dimensional irreducible representation $\overline{\mathrm{PA}}_7$ \cite{Aroyo2011-cr,Aroyo2006-bi1,Aroyo2006-bi2,Elcoro2021-ma,Xu2020-hi} as
\begin{align}
    D^{(\overline{\mathrm{PA}}_7)}(\{C_{2z} | 0\frac{1}{2}0\}) & =
    \begin{pmatrix}
        i & 0 & 0  \\
        0 & i & 0  \\
        0 & 0 & -i
    \end{pmatrix},\label{eq:dkpc2z}                                \\
    D^{(\overline{\mathrm{PA}}_7)}(\{C_{2y} | \frac{1}{2}00\}) & =
    \begin{pmatrix}
        i & 0  & 0 \\
        0 & -i & 0 \\
        0 & 0  & i
    \end{pmatrix},\label{eq:dkpc2y}                                \\
    D^{(\overline{\mathrm{PA}}_7)}(\{C_{3,111}^+ | 000\})      & =
    \begin{pmatrix}
        0 & 0 & -1 \\
        1 & 0 & 0  \\
        0 & 1 & 0
    \end{pmatrix}.\label{eq:dkpc3111}
\end{align}
Using Eqs. \eqref{eq:dkpc2z}, \eqref{eq:dkpc2y}, and \eqref{eq:dkpc3111}, the action of a generator $g = \{ \alpha | \bm b \} \in G_{199}^{\bm k_{\mathrm P}}$ on the $3 \times 3$ Hermitian matrices $S_i$ and the momentum-dependent functions $f_i(\bm k)$ is described as follows
\begin{align}
     & S_i \to D^{(\overline{\mathrm{PA}}_7)}(g)^{\dagger} S_i D^{(\overline{\mathrm{PA}}_7)}(g), \\
     & f_i(\bm k) \to f_i(\alpha \bm k).
\end{align}
By utilizing these transformations, $S_i$ and $f_i(\bm k)$ are classified according to the bases of the irreducible representations of point group T, which constitutes the rotational part of the little group $G_{199}^{\bm k_{\mathrm P}}$. The results of this classification are presented in Table~\ref{tab:T_kBS}. The basis $\bm T = (T_x,T_y,T_z)^{\top}$ for the T representation in the table is chosen such that for the generators
\begin{align}
     & C_{2z}: \bm T \to \begin{pmatrix} -1&0&0\\0&-1&0\\0&0&1 \end{pmatrix} \bm T,      \\
     & C_{2y}: \bm T \to \begin{pmatrix} -1&0&0\\0&1&0\\0&0&-1  \end{pmatrix} \bm T,     \\
     & C_{3,111}^+: \bm T \to \begin{pmatrix} 0&0&1\\1&0&0\\0&1&0   \end{pmatrix} \bm T.
\end{align}
Consequently, the inner product $\bm T \cdot \bm T' = T_x T'_x + T_y T'_y + T_z T'_z$ forms a totally symmetric representation, which can appear in the Hamiltonian. Thus, the $\bm k \cdot \bm p$ Hamiltonian for the threefold fermion is expressed as
\begin{align}
    H_{199,\mathrm P}(\bm k) & = \bm k \cdot (v_1 \bm S_1^{T} + v_2 \bm S_2^{T}),
    \label{eq:H199P}
\end{align}
where $v_1$ and $v_2$ are real coefficients, and $\bm S_1^T = ( S_{1x}^T, S_{1y}^T, S_{1z}^T )$ and $\bm S_2^T= ( S_{2x}^T, S_{2y}^T, S_{2z}^T )$ are basis matrices.
For an applied magnetic field $\bm B$, a similar procedure yields the Zeeman Hamiltonian
\begin{align}
    H_{199,\mathrm P}^Z(\bm B) & = \mu_{\mathrm B} \bm B \cdot (g_1 \bm S_1^{T} + g_2 \bm S_2^{T}),
    \label{eq:H199PZ}
\end{align}
where $g_1$ and $g_2$ are real coefficients.

%%%%%%%%%%%%%%%%%%%%%%%%%%%%%%%%%%%%%%% Subsection %%%%%%%%%%%%%%%%%%%%%%%%%%%%%%%%%%%%%%%

\subsection{Threefold fermion at \texorpdfstring{$\mathrm{\overline P}$}{P-bar} point}
\label{sec:kp_3fold_bar}

The $\overline{\mathrm P}$ point, $\mathrm{\overline P} = (-\pi, -\pi, -\pi)$, is the time-reversal partner of P.
Let $H_{199,\mathrm{\overline P}}(\bm k)$ be the Hamiltonian for the threefold fermion at the $\mathrm{\overline P}$ point, and $H_{199,\mathrm{\overline P}}^{\mathrm Z}(\bm B)$ represent its Zeeman term.
The whole Hamiltonian $H_{199,\mathrm{P \oplus \overline P}}(\bm k, \bm B) = (H_{199,\mathrm P}(\bm k) + H_{199,\mathrm P}^{\mathrm Z}(\bm B)) \oplus (H_{199,\mathrm{\overline P}}(\bm k) + H_{199,\mathrm{\overline P}}^{\mathrm Z}(\bm B))$ must satisfy time-reversal symmetry $\Theta H_{199,\mathrm{P \oplus \overline P}}(\bm k, \bm B) \Theta^\dagger = H_{199,\mathrm{P \oplus \overline P}}(-\bm k, -\bm B)$ under the time-reversal operation $\Theta$ given by \cite{Aroyo2011-cr,Aroyo2006-bi1,Aroyo2006-bi2,Elcoro2021-ma,Xu2020-hi}
\begin{align}
    \Theta = \begin{pmatrix} 0 & -I_3 \\ I_3 & 0 \end{pmatrix} K.
\end{align}
Here, $K$ denotes the complex conjugation operator, and $I_3$ is the $3 \times 3$ identity matrix. Consequently, the Hamiltonian for the threefold fermion at the $\mathrm{\overline P}$ point is expressed as
\begin{align}
    H_{199,\mathrm{\overline P}}(\bm k)             & = \bm k \cdot (- v_1 \bm S_1^{T} + v_2 \bm S_2^{T}),                 \\
    H_{199,\mathrm{\overline P}}^{\mathrm Z}(\bm B) & = \mu_{\mathrm B} \bm B \cdot (- g_1 \bm S_1^{T} + g_2 \bm S_2^{T}).
\end{align}

%%%%%%%%%%%%%%%%%%%%%%%%%%%%%%%%%%%%%%%%%%%%%%%%%%%%%%%%%%%%%%%%%%%%%%%%%%%%%%%%%%%%%%%%%
%%%%%%%%%%%%%%%%%%%%%%%%%%%%%%%%%%%%%%%% Section %%%%%%%%%%%%%%%%%%%%%%%%%%%%%%%%%%%%%%%%
%%%%%%%%%%%%%%%%%%%%%%%%%%%%%%%%%%%%%%%%%%%%%%%%%%%%%%%%%%%%%%%%%%%%%%%%%%%%%%%%%%%%%%%%%

\section{\texorpdfstring{$\bm k \cdot \bm p$}{k.p} model for eightfold fermions under magnetic field}
\label{sec:kp_8fold}

The eightfold fermions emerge at the R point, $\bm k_{\mathrm R} = (\pi,\pi,\pi)$, within SG 218 ($G_{218}$). The little group $G_{218}^{\bm k_{\mathrm R}} \subset G_{218}$ at this point hosts a double-valued eight-dimensional irreducible representation $\overline{\mathrm{R}}_8 \overline{\mathrm{R}}_8$. The representations of the generators of $G_{218}^{\bm k_{\mathrm R}}$ are provided by \cite{Aroyo2011-cr,Aroyo2006-bi1,Aroyo2006-bi2,Elcoro2021-ma,Xu2020-hi} as
\begin{align}
    D^{(\overline{\mathrm{R}}_8\overline{\mathrm{R}}_8)}(\{C_{2z}|000\})                                           & =
    \begin{pmatrix}
        -i\sigma_z &           &           &            \\
                   & i\sigma_z &           &            \\
                   &           & i\sigma_z &            \\
                   &           &           & -i\sigma_z
    \end{pmatrix},                                                                    \\
    D^{(\overline{\mathrm{R}}_8\overline{\mathrm{R}}_8)}(\{C_{2y}|000\})                                           & =
    \begin{pmatrix}
        i\sigma_y &            &           &            \\
                  & -i\sigma_x &           &            \\
                  &            & i\sigma_y &            \\
                  &            &           & -i\sigma_x
    \end{pmatrix},                                                                    \\
    D^{(\overline{\mathrm{R}}_8\overline{\mathrm{R}}_8)}(\{C_{3,111}^+|000\})                                      & =
    \begin{pmatrix}
        s &             &     &          \\
          & s^{\dagger} &     &          \\
          &             & s^* &          \\
          &             &     & s^{\top}
    \end{pmatrix},\label{eq:dkrc3111}                                                                                  \\
    D^{(\overline{\mathrm{R}}_8\overline{\mathrm{R}}_8)}(\{\sigma_{1\bar{1}0}|\frac{1}{2}\frac{1}{2}\frac{1}{2}\}) & =
    \begin{pmatrix}
            & I_2 &     &     \\
        I_2 &     &     &     \\
            &     &     & I_2 \\
            &     & I_2 &
    \end{pmatrix},
\end{align}
where $I_2$ is the $2 \times 2$ identity matrix, and $s$ is defined as
\begin{align}
    s=\frac{1}{\sqrt{2}}
    \begin{pmatrix}
        e^{i5\pi/12} & e^{-i\pi/12}  \\
        e^{i5\pi/12} & e^{i11\pi/12}
    \end{pmatrix}.
\end{align}
Furthermore, the R point is the TRIM. The time-reversal operation \cite{Aroyo2011-cr,Aroyo2006-bi1,Aroyo2006-bi2,Elcoro2021-ma,Xu2020-hi} is given by
\begin{align}
    \Theta = \begin{pmatrix} 0 & -I_4 \\ I_4 & 0 \end{pmatrix} K,
\end{align}
where $I_4$ is the $4 \times 4$ identity matrix. Using these representations, the $8 \times 8$ Hermitian matrices $S_i,(i=1, \ldots, 64)$ and the momentum-dependent functions $f_i(\bm k)$ and magnetic field-dependent functions $f_i(\bm B)$ are classified according to the bases of the irreducible representations of the point group $\mathrm T_d$, which constitutes the rotational part of the little group $G_{218}^{\bm k_{\mathrm R}}$. The results of this classification are presented in Section~\ref{sec:Td_basis}. For the $T_1$ and $T_2$ representations in the tables, the bases $\bm T_1 = (T_{1x},T_{1y},T_{1z})^{\top}$ and $\bm T_2 = (T_{2x},T_{2y},T_{2z})^{\top}$ are chosen to transform under the generators as follows
\begin{align}
     & C_{2z}: \bm T_{1,2} \to \begin{pmatrix} -1 & 0 & 0 \\ 0 & -1 & 0 \\ 0 & 0 & 1 \end{pmatrix} \bm T_{1,2},      \\
     & C_{2y}: \bm T_{1,2} \to \begin{pmatrix} -1 & 0 & 0 \\ 0 & 1 & 0 \\ 0 & 0 & -1 \end{pmatrix} \bm T_{1,2},      \\
     & C_{3,111}^+: \bm T_{1,2} \to \begin{pmatrix} 0 & 0 & 1 \\ 1 & 0 & 0 \\ 0 & 1 & 0 \end{pmatrix} \bm T_{1,2},   \\
     & \sigma_{1\bar{1}0}: \bm T_1 \to \begin{pmatrix} 0 & 1 & 0 \\ 1 & 0 & 0 \\ 0 & 0 & 1 \end{pmatrix} \bm T_1,    \\
     & \sigma_{1\bar{1}0}: \bm T_2 \to \begin{pmatrix} 0 & -1 & 0 \\ -1 & 0 & 0 \\ 0 & 0 & -1 \end{pmatrix} \bm T_2.
\end{align}
Consequently, the products $\bm T_1 \cdot \bm T'_1 = T_{1x} T'_{1x} + T_{1y} T'_{1y} + T_{1z} T'_{1z}$ and $\bm T_2 \cdot \bm T'_2 = T_{2x} T'_{2x} + T_{2y} T'_{2y} + T_{2z} T'_{2z}$ form totally symmetric representations, which can thus appear in the Hamiltonian. Since momentum and magnetic field belong to bases that are odd under time-reversal symmetry for the $T_1$ and $T_2$ representations, the $8 \times 8$ matrices $S_i$ are also chosen as bases odd under time-reversal. Therefore, the $\bm k \cdot \bm p$ Hamiltonian for the eightfold fermion is given by
\begin{align}
    H_{218}(\bm k)
    = \sum_{\alpha=1}^6 v_\alpha \boldsymbol{k} \cdot \boldsymbol{S}_{\alpha}^{T_2^-},
\end{align}
where $v_\alpha$ is a real coefficient. The Zeeman term is expressed as
\begin{align}
    H_{218}^{\mathrm{Z}}(\bm B)
    = \mu_B \sum_{\alpha=1}^4 g_\alpha \boldsymbol{B} \cdot \boldsymbol{S}_{\alpha}^{T_1^-},
\end{align}
where $g_\alpha$ is a real coefficient.

%%%%%%%%%%%%%%%%%%%%%%%%%%%%%%%%%%%%%%%%%%%%%%%%%%%%%%%%%%%%%%%%%%%%%%%%%%%%%%%%%%%%%%%%%
%%%%%%%%%%%%%%%%%%%%%%%%%%%%%%%%%%%%%%%% Section %%%%%%%%%%%%%%%%%%%%%%%%%%%%%%%%%%%%%%%%
%%%%%%%%%%%%%%%%%%%%%%%%%%%%%%%%%%%%%%%%%%%%%%%%%%%%%%%%%%%%%%%%%%%%%%%%%%%%%%%%%%%%%%%%%

\section{Tight-binding model under magnetic field}
\label{sec:3fold_tb}

This section details the construction of a tight-binding model that respects the symmetry of SG 199. The initial step involves constructing a lattice invariant under space group operations by placing lattice points at Wyckoff positions. In this paper, we adopt Wyckoff position 8a of SG 199. The Bravais lattice is body-centered cubic, and its sublattices, expressed in fractional coordinates within the conventional cell \cite{Aroyo2011-cr,Aroyo2006-bi1,Aroyo2006-bi2}, are given by
\begin{align}
    \begin{split}
        \bm n_{\mathrm{A}} & = \ab(x,x,x),               \\
        \bm n_{\mathrm{B}} & = \ab(-x,-x+\frac{1}{2},x), \\
        \bm n_{\mathrm{C}} & = \ab(-x+\frac{1}{2},x,-x), \\
        \bm n_{\mathrm{D}} & = \ab(x,-x,-x+\frac{1}{2}),
    \end{split}
    \label{eq:sublattices}
\end{align}
These four sublattices, along with their eight counterparts generated by centering them with $\bm \tau = (1/2,1/2,1/2)$, define the full set of sublattices. Here, $x$ satisfies $0 \leq x < 1$. Furthermore, the primitive lattice vectors are
\begin{align}
    \begin{split}
        \bm{a}_1 & = \left( -\frac{1}{2},\frac{1}{2},\frac{1}{2} \right), \\
        \bm{a}_2 & = \left( \frac{1}{2},-\frac{1}{2},\frac{1}{2} \right), \\
        \bm{a}_3 & = \left( \frac{1}{2},\frac{1}{2},-\frac{1}{2} \right).
    \end{split}
    \label{eq:primitive_vecs}
\end{align}
The lattice constant is set to 1. This lattice configuration is illustrated in Fig.~\ref{fig:tb199}(a).

The Hamiltonian must be totally symmetric, therefore, is given as
\begin{align}
     & H = \sum_{\bm{k}} \bm c_{\bm{k}}^\dagger H_{\bm{k}} \bm c_{\bm{k}},                                                                \\
     & H_{\bm{k}} = \sum_{\Gamma} \bm t_{\bm{k}}^\Gamma \cdot \bm \sigma^\Gamma,
    \label{eq:Hk}                                                                                                                         \\
     & \ab(\boldsymbol{t}_{\boldsymbol{k}}^\Gamma)_{XY} = \ab(\boldsymbol{t}^\Gamma)_{XY} e^{i \boldsymbol{k} \cdot \boldsymbol{d}_{XY}}.
\end{align}
We define the row vector $\bm c_{\boldsymbol{k}}^\dagger$ by arranging the fermion operators $c_{\boldsymbol{k} \mathrm X, i}^\dagger$, where $c_{\boldsymbol{k} \mathrm X, i}^\dagger$ creates an electron with spin $i$ at sublattice $X$. The real-space fermion operators $c_{\bm m + \bm n_{\mathrm{X}}, i}^\dagger$ are related to the momentum-space operators $c_{\bm k \mathrm{X}, i}^\dagger$ via the Fourier transform
\begin{align}
    c_{\bm k \mathrm{X}, i}^\dagger = \frac{1}{\sqrt{N}} \sum_{\bm m} e^{i \bm k \cdot (\bm m + \bm n_{\mathrm{X}})} c_{\bm m + \bm n_{\mathrm{X}}, i}^\dagger,
\end{align}
where the summation is taken over all primitive unit cells. Note that $\bm t_{\bm k}^\Gamma=\bm t_{\bm k}^{\Gamma \dagger}$, where $\Gamma$ denotes the irreducible representation of point group T, which is compatible with space group 199. The vectors $\bm t^\Gamma = (t_1^\Gamma, \cdots, t_{d}^\Gamma)^\top$ and $\bm \sigma^\Gamma = (\sigma_1^\Gamma, \cdots, \sigma_d^\Gamma)^\top$ correspond to the basis of the $d$-dimensional irreducible representation $\Gamma$ for the hopping and the spin, respectively. The term $t_{\mathrm{XX}}$ represents the onsite energy of sublattice X, while $t_{\mathrm{XY}}$ represents the nearest-neighbor hopping between sublattices X and Y. The matrices $\sigma_0$, $\sigma_x$, $\sigma_y$, and $\sigma_z$ are the identity and Pauli matrices for the spin degrees of freedom, respectively. The vector $\bm d_{\mathrm{XY}}$ points from sublattice X to its nearest-neighbor Y sublattice and is defined as
\begin{align}
    \begin{split}
        \bm d_{\mathrm{XX}} & = \boldsymbol{0},                                    \\
        \bm d_{\mathrm{AB}} & = \bm n_{\mathrm B} - \bm n_{\mathrm A},             \\
        \bm d_{\mathrm{AC}} & = \bm n_{\mathrm C} - \bm n_{\mathrm A},             \\
        \bm d_{\mathrm{AD}} & = \bm n_{\mathrm D} - \bm n_{\mathrm A},             \\
        \bm d_{\mathrm{BC}} & = (\bm n_{\mathrm C} + \bm a_1) - \bm n_{\mathrm B}, \\
        \bm d_{\mathrm{BD}} & = \bm n_{\mathrm D} - (\bm n_{\mathrm B} + \bm a_2), \\
        \bm d_{\mathrm{CD}} & = (\bm n_{\mathrm D} + \bm a_3) - \bm n_{\mathrm C},
    \end{split}
\end{align}

In the following, we derive the transformations of $t$ and $\sigma$. For a generator $g=\{\alpha | \bm b\} \in G_{199}$, the fermion operator transforms as $g:c_{\bm k}^\dagger \to  c_{\alpha\bm k}^\dagger D_{\boldsymbol{k}}(g)$, where
\begin{align}
    \ab(D_{\boldsymbol{k}}(g))_{X'i', Xi} = D^{\mathrm{sl}}_{\mathrm{X'X}}(g) D^{\mathrm{sp}}_{i'i}(\alpha) e^{- i \alpha \bm k \cdot \bm b}.
    \label{eq:Dk}
\end{align}
Here, $D^{\mathrm{sl}}(g)$ is the permutation matrix representing the sublattice (sl) transformation, and $D^{\mathrm{sp}}(\alpha)$ is the matrix representing the spin (sp) space rotation. Specifically, for the generators of SG 199, in the basis of (A,B,C,D), they are
\begin{align}
    D^{\mathrm{sl}}(\{C_{2z} | \frac{1}{2}0\frac{1}{2}\}) & = \begin{pmatrix} 0 & 1 & 0 & 0 \\ 1 & 0 & 0 & 0 \\ 0 & 0 & 0 & 1 \\ 0 & 0 & 1 & 0 \end{pmatrix}, \\
    D^{\mathrm{sl}}(\{C_{2y} | 0\frac{1}{2}\frac{1}{2}\}) & = \begin{pmatrix} 0 & 0 & 1 & 0 \\ 0 & 0 & 0 & 1 \\ 1 & 0 & 0 & 0 \\ 0 & 1 & 0 & 0 \end{pmatrix}, \\
    D^{\mathrm{sl}}(\{C^+_{3,111} | 000\})                & = \begin{pmatrix}
                                                                  1 & 0 & 0 & 0 \\
                                                                  0 & 0 & 1 & 0 \\
                                                                  0 & 0 & 0 & 1 \\
                                                                  0 & 1 & 0 & 0\end{pmatrix}.
\end{align}
From the above, Eq.~\eqref{eq:Hk} transforms as
\begin{align}
    g: H_{\boldsymbol{k}}
     & \to D_{\alpha^{-1}\boldsymbol k}(g)
    H_{\alpha^{-1}\boldsymbol{k}}
    D_{\alpha^{-1}\boldsymbol{k}}(g)^\dagger
    =
    \sum_{\Gamma}
    \boldsymbol{t}_{\boldsymbol{k}}^{\Gamma\prime}
    \cdot
    \boldsymbol{\sigma}^{\Gamma\prime}.
\end{align}
$t$ and $\sigma$ are transformed as
\begin{align}
    \ab(t_{\boldsymbol{k} a}^{\Gamma\prime})_{XY}
     & =
    D^{\mathrm{sl}}(g) \ab(t_{a}^\Gamma)_{XY} D^{\mathrm{sl}}(g)^\dag e^{i\boldsymbol{k} \cdot \boldsymbol{d}_{XY}},
    \\
    \sigma_a^{\Gamma\prime}
     & =
    D^{\mathrm{sp}}(\alpha) \sigma_a^\Gamma D^{\mathrm{sp}}(\alpha)^\dag.
\end{align}
where the following relation holds
\begin{align}
    \sum_{X'Y'}
    D^{\mathrm{sl}}_{XX'}(g)
    e^{i \alpha^{-1} \boldsymbol{k} \cdot \boldsymbol{d}_{X'Y'}}
    D_{YY'}^{\mathrm{sl}}(g)^*
    =
    e^{i \boldsymbol{k} \cdot \boldsymbol{d}_{XY}}.
\end{align}
We can verify that the Hamiltonian is totally symmetric as $D_{\boldsymbol{k}}(g) H_{\boldsymbol{k}} D_{\boldsymbol{k}}(g)^\dag = H_{\alpha \boldsymbol{k}}$ since $\boldsymbol{t}^\Gamma$ and $\boldsymbol{\sigma}^\Gamma$ are irreps $\Gamma$ for the sublattice and spin, by definition, then $\boldsymbol{t}^{\Gamma\prime} = D^\Gamma(g) \boldsymbol{t}^\Gamma$ and $\boldsymbol{\sigma}^{\Gamma\prime} = D^\Gamma(g) \boldsymbol{\sigma}^\Gamma$ hold, where $D^\Gamma(g)$ denotes the irreducible representation $\Gamma$ of $g$.

As a result, we find $t_a^\Gamma$ and $\sigma_a^\Gamma$ by using the projection onto $a$th basis of irrep $\Gamma$
\begin{align}
    P^{\Gamma}_{\alpha} = & \frac{d_{\Gamma}}{|\mathrm T|} \sum_{g \in \mathrm T} D_{\alpha\alpha}^{\Gamma}(g)^* g.
\end{align}
Here, $d_{\Gamma}$ is the dimension of the representation, and $|\mathrm T|$ is the order of the group. By utilizing projection operators, the hopping $\bm t$ and spin $\bm \sigma$ are classified according to the irreducible representations of point group T, as presented in Table~\ref{tab:T_ts}. For simplicity, the set of onsite energies are denoted by
\begin{align}
    \boldsymbol{\epsilon} = (t_{\mathrm{AA}}, t_{\mathrm{BB}}, t_{\mathrm{CC}}, t_{\mathrm{DD}}),
\end{align}
and the set of nearest-neighbor hopping parameters by
\begin{align}
    \bm t^{\Gamma} = ( t_{\mathrm{AB}},t_{\mathrm{AC}},t_{\mathrm{AD}},t_{\mathrm{BC}},t_{\mathrm{BD}},t_{\mathrm{CD}} ).
\end{align}
Furthermore, the basis $\bm T = (T_x,T_y,T_z)^{\top}$ for the T representation in the table is chosen such that for the generators
\begin{align}
     & C_{2z}: \bm T \to \begin{pmatrix} -1&0&0\\0&-1&0\\0&0&1 \end{pmatrix} \bm T,  \\
     & C_{2y}: \bm T \to \begin{pmatrix} -1&0&0\\0&1&0\\0&0&-1  \end{pmatrix} \bm T, \\
     & C_{3,111}^+: \bm T \to \begin{pmatrix}
                                  0 & 1 & 0 \\
                                  0 & 0 & 1 \\
                                  1 & 0 & 0\end{pmatrix} \bm T.
\end{align}

The onsite energies are decomposed into the irreducible representations $A+T$. The contribution of $A$, $\boldsymbol{\epsilon}^{A^+}$, is an onsite term that takes the same value at every site; it can be removed by redefining the zero energy. The $T$ components appear in the Hamiltonian since $\boldsymbol{T} \cdot \boldsymbol{T'}$ is totally symmetric. However, since $\boldsymbol{\epsilon}^{T^+}$ is real and even under time reversal, whereas both spin and the magnetic field are odd, the terms $\boldsymbol{\epsilon}^{T^+} \cdot \boldsymbol{\sigma}$ and $\boldsymbol{\epsilon}^{T^+} \cdot \boldsymbol{B}$, although they transform as the $A$ representation, are forbidden in the Hamiltonian by time-reversal symmetry.

The inner product $\bm T \cdot \bm T' = T_x T'_x + T_y T'_y + T_z T'_z$ forms a totally symmetric representation, which can appear in the Hamiltonian. By constructing a totally symmetric representation from the product representation of the hopping and spin, the tight-binding model invariant under SG 199 symmetry is given as
\begin{align}
    H_{199\boldsymbol{k}}
    =
    \lambda_1 t_{\boldsymbol{k}}^{A+}
    +
    \lambda_2 \boldsymbol{t}_{1 \boldsymbol{k}}^{T-} \cdot \boldsymbol{\sigma}
    +
    \lambda_3 \boldsymbol{t}_{2 \boldsymbol{k}}^{T-} \cdot \boldsymbol{\sigma}.
    \label{eq:H199k}
\end{align}
Additionally, the Zeeman term, when a magnetic field is applied, is given by
\begin{align}
    H_{199 \boldsymbol{k}}^{\mathrm Z}(\bm B) & = \mu_{\mathrm{B}} \bm B \cdot \bm \sigma.
\end{align}

%%%%%%%%%%%%%%%%%%%%%%%%%%%%%%%%%%%%%%%%%%%%%%%%%%%%%%%%%%%%%%%%%%%%%%%%%%%%%%%%%%%%%%%%%
%%%%%%%%%%%%%%%%%%%%%%%%%%%%%%%%%%%%%%%% Section %%%%%%%%%%%%%%%%%%%%%%%%%%%%%%%%%%%%%%%%
%%%%%%%%%%%%%%%%%%%%%%%%%%%%%%%%%%%%%%%%%%%%%%%%%%%%%%%%%%%%%%%%%%%%%%%%%%%%%%%%%%%%%%%%%

\section{Correspondence between the \texorpdfstring{$\bm k \cdot \bm p$}{k.p} and Tight-Binding models for threefold fermions}
\label{sec:3fold_correspondence}

This section establishes the correspondence between the parameters of the $\bm k \cdot \bm p$ model $(v_1, v_2, g_1, g_2)$ and those of the tight-binding model $(\lambda_1, \lambda_2, \lambda_3)$ for the threefold fermion. This is achieved by constructing an effective model from the tight-binding Hamiltonian in the vicinity of the P point. We begin by classifying the basis, which is composed of the sublattice and spin degrees of freedom within the tight-binding model, according to the irreducible representations of the little group $G^{\bm k_{\mathrm P}}_{199}$. The sublattice basis constitutes the single-value representations $\overline{\mathrm{PA}}_1$ and $\overline{\mathrm{PA}}_3$ \cite{Aroyo2011-cr,Aroyo2006-bi1,Aroyo2006-bi2,Elcoro2021-ma,Xu2020-hi} as
\begin{align}
     & D^{(\mathrm{PA_1})}(\{C_{2z}|0\frac{1}{2}0\}) = D^{(\mathrm{PA_3})}(\{C_{2z}|0\frac{1}{2}0\})  =
    \begin{pmatrix}
        1 & 0  \\
        0 & -1
    \end{pmatrix},                                                                                      \\
     & D^{(\mathrm{PA_1})}(\{C_{2y}|\frac{1}{2}00\}) = D^{(\mathrm{PA_3})}(\{C_{2y}|\frac{1}{2}00\})  =
    \begin{pmatrix}
        0 & 1 \\
        1 & 0
    \end{pmatrix},                                                                                      \\
     & D^{(\mathrm{PA_1})}(\{C^+_{3,111}|000\})       = \frac{1}{\sqrt{2}}
    \begin{pmatrix}
        e^{-i \pi /12} & e^{-i\pi/12}   \\
        e^{i 5\pi/12}  & e^{-i 7\pi/12}
    \end{pmatrix},                                                                     \\
     & D^{(\mathrm{PA_3})}(\{C^+_{3,111}|000\})       = \frac{1}{\sqrt{2}}
    \begin{pmatrix}
        e^{i 7\pi/12}   & e^{i7\pi/12} \\
        e^{-i 11\pi/12} & e^{i \pi/12}
    \end{pmatrix}.
\end{align}
These basis states are explicitly constructed from the sublattice basis operators $\bm c_{\bm k X, \sigma}^\dagger$ as follows
\begin{align}
    (c_{\bm k \sigma}^{(\mathrm{PA_1}1) \dagger}, c_{\bm k \sigma}^{(\mathrm{PA_1}2) \dagger}) = \bm c_{\bm k X, \sigma}^\dagger U_1, \\
    (c_{\bm k \sigma}^{(\mathrm{PA_3}1) \dagger}, c_{\bm k \sigma}^{(\mathrm{PA_3}2) \dagger}) = \bm c_{\bm k X, \sigma}^\dagger U_2,
\end{align}
where $\bm c_{\bm k X, \sigma}^\dagger = (c_{\bm k \mathrm A, \sigma}^\dagger, c_{\bm k \mathrm B, \sigma}^\dagger, c_{\bm k \mathrm C, \sigma}^\dagger, c_{\bm k \mathrm D, \sigma}^\dagger)$. The matrices $U_1$ and $U_2$ are defined as follows
\begin{align}
    U_1 & = \frac{1}{\sqrt{6 - 2\sqrt{3}}} \begin{pmatrix} 1 & u_1^* \\ -ie^{4i\pi x} & u_1e^{4i\pi x} \\ -u_1e^{4i\pi x} & -ie^{4i\pi x} \\ u_1^*e^{4i\pi x} & -e^{4i\pi x} \end{pmatrix},   \\
    U_2 & = \frac{1}{\sqrt{6 + 2\sqrt{3}}} \begin{pmatrix} 1 & -u_2^* \\ -ie^{4i\pi x} & -u_2e^{4i\pi x} \\ u_2e^{4i\pi x} & -ie^{4i\pi x} \\ -u_2^*e^{4i\pi x} & -e^{4i\pi x} \end{pmatrix},
\end{align}
where $u_1 = (1+i)(-1+\sqrt{3})/2$, $u_2=(1+i)(1+\sqrt{3})/2$, and $x$ follows its definition in Eq.~\eqref{eq:sublattices}.

From these single-value representations, it is possible to construct the basis for the double-value representation $\overline{\mathrm{PA}}_7$ of the little group $G^{\bm k_{\mathrm P}}_{199}$, Eqs.~\eqref{eq:dkpc2z}--\eqref{eq:dkpc3111}, which protects the threefold fermion. This basis consists of two sets $\bm c_{1\bm k}^{(\overline{\mathrm{PA}}_7) \dagger}=(c_{1\bm k}^{(\overline{\mathrm{PA}}_7 1) \dagger},c_{1\bm k}^{(\overline{\mathrm{PA}}_7 2) \dagger},c_{1\bm k}^{(\overline{\mathrm{PA}}_7 3) \dagger })$ and $\bm c_{2\bm k}^{(\overline{\mathrm{PA}}_7) \dagger}=(c_{2\bm k}^{(\overline{\mathrm{PA}}_7 1) \dagger},c_{2\bm k}^{(\overline{\mathrm{PA}}_7 2) \dagger},c_{2\bm k}^{(\overline{\mathrm{PA}}_7 3) \dagger })$. The components of these representations are explicitly defined as
\begin{align}
    c_{1\bm k}^{(\overline{\mathrm{PA}}_7 1) \dagger} & = \frac{1}{\sqrt{2}}( c_{\bm k \downarrow}^{(\mathrm{PA_1}1) \dagger} + i c_{\bm k \uparrow}^{(\mathrm{PA_1}2) \dagger}),          \\
    c_{1\bm k}^{(\overline{\mathrm{PA}}_7 2) \dagger} & = \frac{(-1)^{1/6}}{\sqrt{2}}(c_{\bm k \downarrow}^{(\mathrm{PA_1}1) \dagger} - i c_{\bm k \uparrow}^{(\mathrm{PA_1}2) \dagger}),  \\
    c_{1\bm k}^{(\overline{\mathrm{PA}}_7 3) \dagger} & = \frac{(-1)^{4/3}}{\sqrt{2}} (c_{\bm k \uparrow}^{(\mathrm{PA_1}1) \dagger} - i c_{\bm k \downarrow}^{(\mathrm{PA_1}2) \dagger}), \\
    c_{2\bm k}^{(\overline{\mathrm{PA}}_7 1) \dagger} & = \frac{1}{\sqrt{2}}( c_{\bm k \downarrow}^{(\mathrm{PA_3}1) \dagger} + i c_{\bm k \uparrow}^{(\mathrm{PA_3}2) \dagger}),          \\
    c_{2\bm k}^{(\overline{\mathrm{PA}}_7 2) \dagger} & = \frac{(-1)^{1/6}}{\sqrt{2}}(c_{\bm k \downarrow}^{(\mathrm{PA_3}1) \dagger} - i c_{\bm k \uparrow}^{(\mathrm{PA_3}2) \dagger}),  \\
    c_{2\bm k}^{(\overline{\mathrm{PA}}_7 3) \dagger} & = \frac{(-1)^{4/3}}{\sqrt{2}} (c_{\bm k \uparrow}^{(\mathrm{PA_3}1) \dagger} - i c_{\bm k \downarrow}^{(\mathrm{PA_3}2) \dagger}).
\end{align}
We consider the Hamiltonian matrix $H_{199\bm k_{\mathrm P}}$ from Eq.~\eqref{eq:H199k} projected onto the $\overline{\mathrm{PA}}_7$ representation basis, referred to as the band basis, denoted by $H_{199\bm k_{\mathrm P}}^{\mathrm{TF}}$. The band basis is chosen as $\bm c_{\bm k}^{\mathrm{TF}\dagger} = ((a-b) \bm c_{1\bm k}^{(\overline{\mathrm{PA}}_7) \dagger} + \bm c_{2\bm k}^{(\overline{\mathrm{PA}}_7) \dagger}, \, (a+b) \bm c_{1\bm k}^{(\overline{\mathrm{PA}}_7) \dagger} + \bm c_{2\bm k}^{(\overline{\mathrm{PA}}_7) \dagger})$, where $a$ and $b$ are defined as
\begin{align}
    a & = \frac{ -3 \lambda_1 + \lambda_2 - \lambda_3 }{\sqrt{2}(1-i\sqrt{3})\lambda_2 + 2\sqrt{2}\lambda_3},                                                                                     \\
    b & = \frac{ \sqrt{9( \lambda_1^2 + \lambda_2^2 + \lambda_3^2 ) + 6(-\lambda_1 \lambda_2 + \lambda_1 \lambda_3 + \lambda_2 \lambda_3)}}{\sqrt{2}(1-i\sqrt{3})\lambda_2 + 2\sqrt{2}\lambda_3}.
\end{align}
When $H_{199\bm k_{\mathrm P}}$ is projected onto this band basis, the resulting Hamiltonian $H_{199\bm k_{\mathrm P}}^{\mathrm{TF}}$ is diagonal
\begin{align}
    H_{199 \boldsymbol{k}_{\mathrm{P}}}^{\mathrm{TF}}
     &
    =
    \bm c_{\bm k}^{\mathrm{TF}\dagger}
    H_{199\bm k_{\mathrm P}}
    \bm c_{\bm k}^{\mathrm{TF}}
    \notag \\&
    =
    \bm c_{\bm k}^{\mathrm{TF}\dagger}
    \mathrm{diag}(\epsilon_+, \epsilon_+, \epsilon_+, \epsilon_-, \epsilon_-, \epsilon_-)
    \bm c_{\bm k}^{\mathrm{TF}},
\end{align}
where
\begin{align}
    \epsilon_{\pm} = \pm  \sqrt{3( \lambda_1^2 + \lambda_2^2 + \lambda_3^2 ) + 2(-\lambda_1 \lambda_2 + \lambda_1 \lambda_3 + \lambda_2 \lambda_3)}.
\end{align}
represent the two energy levels of the threefold fermions at the P point.
Expanding the tight-binding model around the P point to first order in momentum, we obtain
\begin{align}
    H_{199 \bm k_{\mathrm P} + \delta \bm k}
     &
    = H_{199 \bm k_{\mathrm P}}
    +
    \delta \bm k \cdot \partial_{\bm k} H_{199 \bm k_{\mathrm P}}.
\end{align}
The second term, projected onto the band basis $\bm c_{\bm k}^{\mathrm{TF}}$, as $H_{199 \delta \bm k}^{\mathrm{TF}}$, is given by
\begin{align}
    H_{199 \delta \bm k}^{\mathrm{TF}}
     &
    =
    \boldsymbol{c}_{\boldsymbol{k}}^{\mathrm{TF}^\dag}
    \delta \bm k \cdot \partial_{\bm k} H_{199 \bm k_{\mathrm P}}
    \boldsymbol{c}_{\boldsymbol{k}}^{\mathrm{TF}}
    \notag \\&
    =
    \boldsymbol{c}_{\boldsymbol{k}}^{\mathrm{TF}^\dag}
    \begin{pmatrix}
        H_{199, \mathrm P}^+(\delta \bm k) & C_{\delta \bm k} \\
        C_{\delta \bm k}^\dagger           &
        {H}_{199, \mathrm P}^-(\delta \bm k)
    \end{pmatrix}
    \boldsymbol{c}_{\boldsymbol{k}}^{\mathrm{TF}},
\end{align}
where $H_{199, \mathrm P}^\pm(\delta \bm k) = \sum_{i=1}^2 (v_i^\pm \delta\boldsymbol{k} + g_i^\pm \boldsymbol{B}) \cdot \boldsymbol{S}_i^T$ [see Eqs.~(\ref{eq:H199P}) and \eqref{eq:H199PZ}] represent the effective Hamiltonians for the threefold fermions at energies $\epsilon_{\pm}$, respectively, while $C_{\delta \bm k}$ is the hybridization term between them. From this, the parameters of the $\bm k \cdot \bm p$ Hamiltonian can be expressed as
\begin{align}
    v_1^\pm & = \frac{1}{4} ( -\lambda_1 - \lambda_2 + \lambda_3),                                                                                                                                                                                                       \\
    v_2^\pm & = \pm \frac{ \lambda_1^2 + \lambda_2^2 + \lambda_3^2 + 2(- \lambda_1 \lambda_2 + \lambda_1 \lambda_3 + \lambda_2 \lambda_3)}{4 \sqrt{3( \lambda_1^2 + \lambda_2^2 + \lambda_3^2 ) + 2(-\lambda_1 \lambda_2 + \lambda_1 \lambda_3 + \lambda_2 \lambda_3)}}.
\end{align}
Performing similar calculations for the Zeeman term is determined as
\begin{align}
    g_1^\pm & = \pm \frac{ -3\lambda_1 + \lambda_2 - \lambda_3 }{2 \sqrt{3( \lambda_1^2 + \lambda_2^2 + \lambda_3^2 ) + 2(-\lambda_1 \lambda_2 + \lambda_1 \lambda_3 + \lambda_2 \lambda_3)}}, \\
    g_2^\pm & = \frac{1}{2}.
\end{align}

%%%%%%%%%%%%%%%%%%%%%%%%%%%%%%%%%%%%%%%%%%%%%%%%%%%%%%%%%%%%%%%%%%%%%%%%%%%%%%%%%%%%%%%%%
%%%%%%%%%%%%%%%%%%%%%%%%%%%%%%%%%%%%%%%% Section %%%%%%%%%%%%%%%%%%%%%%%%%%%%%%%%%%%%%%%%
%%%%%%%%%%%%%%%%%%%%%%%%%%%%%%%%%%%%%%%%%%%%%%%%%%%%%%%%%%%%%%%%%%%%%%%%%%%%%%%%%%%%%%%%%

\section{Matrix basis of point group \texorpdfstring{$T_d$}{Td}}
\label{sec:Td_basis}

This section provides a detailed classification of the $8 \times 8$ Hermitian matrices based on the irreducible representations of the point group $\mathrm T_d$. The character table for $\mathrm T_d$, along with the basis of momentum $\bm k$ and magnetic field $\bm B$, is presented in Table~\ref{tab:Td_kb}. The systematic classification of the 64 $8 \times 8$ Hermitian matrices into bases for these irreducible representations is then detailed in Table~\ref{tab_Td_basis_1} and Table~\ref{tab_Td_basis_2}. The $4 \times 4$ matrices $P_i$ (for $i=1, \ldots, 12$) used in these tables are defined as follows
\begin{align}
    \begin{split}
         & P_1 = \begin{pmatrix} \sigma_x & 0 \\ 0 & \sigma_y \end{pmatrix},
        P_2 =\begin{pmatrix} \sigma_y & 0 \\ 0 & \sigma_x \end{pmatrix},
        P_3 =\begin{pmatrix}\sigma_z & 0 \\ 0 & \sigma_z \end{pmatrix},                                   \\
         & P_4 = \begin{pmatrix} \sigma_x & 0 \\ 0 & -\sigma_y \end{pmatrix},
        P_5 = \begin{pmatrix} \sigma_y & 0 \\ 0 & -\sigma_x \end{pmatrix},
        P_6 = \begin{pmatrix}\sigma_z & 0 \\ 0 & -\sigma_z\end{pmatrix},                                  \\
         & P_7 =\begin{pmatrix} 0 & \sigma_x + \sigma_y \\ \sigma_x + \sigma_y & 0 \end{pmatrix},         \\
         & P_8 = \begin{pmatrix} 0 & i(\sigma_x + \sigma_y) \\ -i(\sigma_x+\sigma_y) & 0 \end{pmatrix},   \\
         & P_9 = \begin{pmatrix}\sigma_x & 0 \\ 0 & \sigma_x \end{pmatrix},
        P_{10} = \begin{pmatrix} \sigma_x & 0 \\ 0 & -\sigma_x \end{pmatrix},                             \\
         & P_{11} = \begin{pmatrix} 0 & \sigma_0 + i \sigma_z \\ \sigma_0 + i \sigma_z & 0 \end{pmatrix},
        P_{12} = \begin{pmatrix} 0 & - \sigma_x \\ \sigma_x & 0 \end{pmatrix}.
    \end{split}
\end{align}

\begin{table*}
    \caption{Basis matrices for point group $\mathrm T_d$. $a=e^{i\pi/3},\,a_+=e^{i\pi/3}+1,\,a_-=e^{i\pi/3}-1,\,[A,B]=AB-BA$. $\Gamma$ denotes irreps of the point group $\mathrm T_d$. The third column defines the notation for the $8 \times 8$ matrices corresponding to each irrep. The fourth column provides the explicit $8 \times 8$ matrices for the symbols defined in the third row.}
    \begin{ruledtabular}
        \begin{tabular}{cccc}
            $\Gamma$ & TR  & Symbol                                                              & Definition                                     \\
            \hline
                     &     &                                                                     &                                                \\
            $A_1$    & $+$ & $S^{A_1^+}$                                                         & $I_8$                                          \\
                     &     &                                                                     &                                                \\
                     & $-$ & $S_1^{A_1^-}$                                                       & $
            \begin{pmatrix}
                    0                     & 0                     & 0                     & \sigma_z + i \sigma_0 \\
                    0                     & 0                     & \sigma_z + i \sigma_0 & 0                     \\
                    0                     & \sigma_z - i \sigma_0 & 0                     & 0                     \\
                    \sigma_z - i \sigma_0 & 0                     & 0                     & 0
                \end{pmatrix}$                                         \\
                     &     &                                                                     &                                                \\
                     &     & $S_2^{A_1^-}$                                                       & $
            \begin{pmatrix}
                    0                       & 0                        & 0                     & i \sigma_z - \sigma_0 \\
                    0                       & 0                        & i \sigma_z - \sigma_0 & 0                     \\
                    0                       & - i \sigma_z -  \sigma_0 & 0                     & 0                     \\
                    - i \sigma_z - \sigma_0 & 0                        & 0                     & 0
                \end{pmatrix}$                                    \\
                     &     &                                                                     &                                                \\
            \hline
                     &     &                                                                     &                                                \\
            $A_2$    & $-$ & $S_1^{A_2^-} $                                                      & $
            \begin{pmatrix}
                    0                                & 0                              & 0                               & (i-1)\sigma_z -(i+1) \sigma_0 \\
                    0                                & 0                              & -(i-1)\sigma_z + (i+1) \sigma_0 & 0                             \\
                    0                                & (i+1)\sigma_z + (i-1) \sigma_0 & 0                               & 0                             \\
                    - (i+1)\sigma_z - (i-1) \sigma_0 & 0                              & 0                               & 0
                \end{pmatrix}$   \\
                     &     &                                                                     &                                                \\
                     &     & $ S_2^{A_2^-} $                                                     & $
            \begin{pmatrix}
                    0                                & 0                              & 0                                & (i+1) \sigma_z +(i-1) \sigma_0 \\
                    0                                & 0                              & - (i+1) \sigma_z -(i-1) \sigma_0 & 0                              \\
                    0                                & (i-1) \sigma_z +(i+1) \sigma_0 & 0                                & 0                              \\
                    -(i-1) \sigma_z + (i+1) \sigma_0 & 0                              & 0                                & 0
                \end{pmatrix}$ \\
                     &     &                                                                     &                                                \\
            \hline
                     &     &                                                                     &                                                \\
            $E$      & $+$ & $ [S_{2z^2-x^2-y^2}^{E^+}, S_{x^2-y^2}^{E^+}] $                     & $
                \left[ \begin{pmatrix}
                               0                  & -\sigma_x+\sigma_y & 0                    & 0                    \\
                               -\sigma_x+\sigma_y & 0                  & 0                    & 0                    \\
                               0                  & 0                  & 0                    & -\sigma_x+\sigma_y^* \\
                               0                  & 0                  & -\sigma_x+\sigma_y^* & 0
                           \end{pmatrix},
                    \frac{i}{\sqrt{3}}
            \begin{pmatrix}
                        0                  & \sigma_x-\sigma_y & 0                   & 0                    \\
                        -\sigma_x+\sigma_y & 0                 & 0                   & 0                    \\
                        0                  & 0                 & 0                   & -\sigma_x+\sigma_y^* \\
                        0                  & 0                 & \sigma_x-\sigma_y^* & 0
                    \end{pmatrix}\right] $                                                   \\
                     &     &                                                                     &                                                \\
                     & $-$ & $                [S_{1,2z^2-x^2-y^2}^{E^-} , S_{1,x^2-y^2}^{E^-}] $ & $
                \left[ \begin{pmatrix}
                               0                  & -\sigma_x+\sigma_y & 0                 & 0                 \\
                               -\sigma_x+\sigma_y & 0                  & 0                 & 0                 \\
                               0                  & 0                  & 0                 & \sigma_x+\sigma_y \\
                               0                  & 0                  & \sigma_x+\sigma_y & 0
                           \end{pmatrix},
                    \frac{i}{\sqrt{3}}
            \begin{pmatrix}
                        0                  & \sigma_x-\sigma_y & 0                  & 0                 \\
                        -\sigma_x+\sigma_y & 0                 & 0                  & 0                 \\
                        0                  & 0                 & 0                  & \sigma_x+\sigma_y \\
                        0                  & 0                 & -\sigma_x-\sigma_y & 0
                    \end{pmatrix} \right]$                                                       \\
                     &     &                                                                     &                                                \\
                     &     & $[S_{2,2z^2-x^2-y^2}^{E^-},S_{2,x^2-y^2}^{E^-}]$                    & $
                \left[ \begin{pmatrix}
                               0              & 0            & i a \sigma_y & 0             \\
                               0              & 0            & 0            & ia^* \sigma_y \\
                               -ia^* \sigma_y & 0            & 0            & 0             \\
                               0              & -ia \sigma_y & 0            & 0
                           \end{pmatrix},
            \begin{pmatrix}
                        0             & 0                  & -ia_+^{*} \sigma_y & 0               \\
                        0             & 0                  & 0                  & -i a^+ \sigma_y \\
                        ia^+ \sigma_y & 0                  & 0                  & 0               \\
                        0             & i a_+^{*} \sigma_y & 0                  & 0
                    \end{pmatrix} \right]$                                                             \\
                     &     &                                                                     &                                                \\
                     &     & $[S_{3,2z^2-x^2-y^2}^{E^-},S_{3,x^2-y^2}^{E^-}] $                   & $
                \left[ \begin{pmatrix}
                               0            & 0                & a_-^{*} \sigma_y & 0            \\
                               0            & 0                & 0                & a_- \sigma_y \\
                               a_- \sigma_y & 0                & 0                & 0            \\
                               0            & a_-^{*} \sigma_y & 0                & 0
                           \end{pmatrix},
            \begin{pmatrix}
                        0            & 0                & a_+^{*} \sigma_y & 0            \\
                        0            & 0                & 0                & a_+ \sigma_y \\
                        a_+ \sigma_y & 0                & 0                & 0            \\
                        0            & a_+^{*} \sigma_y & 0                & 0
                    \end{pmatrix} \right]$                                                                     \\
            \label{tab_Td_basis_1}
        \end{tabular}
    \end{ruledtabular}
\end{table*}

\begin{table*}
    \renewcommand{\arraystretch}{1.5}
    \caption{Basis matrices for point group $T_d$. The matrix $Q$ is the 8-dimensional representation of $C_{3,111}^+$ from Eq.~\eqref{eq:dkrc3111}, defined as $Q=D^{\bm k_{\mathrm R}}(\{C^+_{3,111}|000\})$}
    \begin{ruledtabular}
        \begin{tabular}{cccc}
            $\Gamma$ & TR  & Symbol                                                                        & Definition                                                                                                                                                                                                                                                                                                           \\
            \hline
                     &     &                                                                               &                                                                                                                                                                                                                                                                                                                      \\
            $T_1$    & $+$ & $[S_{1,xz^2-xy^2}^{T_1^+}, S_{1,yz^2-yx^2}^{T_1^+}, S_{1,zx^2-zy^2}^{T_1^+}]$ & $ \left[ \begin{pmatrix} P_4 & 0 \\ 0 & P_4^* \end{pmatrix}, \,  \begin{pmatrix} P_5 & 0 \\ 0 & P_5^* \end{pmatrix}, \, \begin{pmatrix} P_6 & 0 \\ 0 & P_6^* \end{pmatrix} \right]$                                                                                                                                  \\
                     &     & $[S_{2,xz^2-xy^2}^{T_1^+}, S_{2,yz^2-yx^2}^{T_1^+}, S_{2,zx^2-zy^2}^{T_1^+}]$ & $ \left[i \,[S_{1,yz^2-yx^2}^{T_1^+}, S_{zx^2-zy^2}^{T_1^+}],\, i \, [ S_{zx^2-zy^2}^{T_1^+}, S_{1,xz^2-xy^2}^{T_1^+}],\,i\, [S_{1,xz^2-xy^2}^{T_1^+}, S_{1,yz^2-yx^2}^{T_1^+}]\right]$                                                                                                                              \\
                     &     & $[S_{3,xz^2-xy^2}^{T_1^+}, S_{3,yz^2-yx^2}^{T_1^+}, S_{3,zx^2-zy^2}^{T_1^+}]$ & $ \left[ Q^{2\dagger} \begin{pmatrix} P_8 & 0 \\ 0 & P_8^* \end{pmatrix} Q^2, \, Q^\dagger \begin{pmatrix} P_8 & 0 \\ 0 & P_8^* \end{pmatrix} Q,\,\begin{pmatrix} P_8 & 0 \\ 0 & P_8^* \end{pmatrix} \right] $                                                                                                       \\
                     &     & $[S_{4,xz^2-xy^2}^{T_1^+}, S_{4,yz^2-yx^2}^{T_1^+}, S_{4,zx^2-zy^2}^{T_1^+}]$ & $ \left[ Q^{2\dagger} \begin{pmatrix} 0 & P_{12} \\ P_{12}^* & 0 \end{pmatrix} Q^2,\,Q^\dagger \begin{pmatrix} 0 & P_{12} \\ P_{12}^* & 0 \end{pmatrix} Q, \, \begin{pmatrix} 0 & P_{12} \\ P_{12}^* & 0 \end{pmatrix} \right]$                                                                                      \\
                     &     & $[S_{5,xz^2-xy^2}^{T_1^+}, S_{5,yz^2-yx^2}^{T_1^+}, S_{5,zx^2-zy^2}^{T_1^+}]$ & $ \left[ Q^{2\dagger} \begin{pmatrix} 0 & i P_{12} \\ - i P_{12}^* & 0 \end{pmatrix} Q^2,\, Q^\dagger \begin{pmatrix} 0 & i P_{12} \\ - i P_{12}^* & 0 \end{pmatrix} Q, \, \begin{pmatrix} 0 & i P_{12} \\ - i P_{12}^* & 0 \end{pmatrix} \right] $                                                                  \\
                     & $-$ & $[S_{1,xz^2-xy^2}^{T_1^-}, S_{1,yz^2-yx^2}^{T_1^-}, S_{1,zx^2-zy^2}^{T_1^-}]$ & $ \left[ \begin{pmatrix} P_4 & 0 \\ 0 & -P_4^* \end{pmatrix}, \,  \begin{pmatrix} P_5 & 0 \\ 0 & -P_5^* \end{pmatrix}, \, \begin{pmatrix} P_6 & 0 \\ 0 & -P_6^* \end{pmatrix} \right]$                                                                                                                               \\
                     &     & $[S_{2,xz^2-xy^2}^{T_1^-}, S_{2,yz^2-yx^2}^{T_1^-}, S_{2,zx^2-zy^2}^{T_1^-}]$ & $ \left[ Q^{2\dagger} \begin{pmatrix} P_8 & 0 \\ 0 & -P_8^* \end{pmatrix} Q^2, \, Q^\dagger \begin{pmatrix} P_8 & 0 \\ 0 & -P_8^* \end{pmatrix} Q,\, \begin{pmatrix} P_8 & 0 \\ 0 & -P_8^* \end{pmatrix} \right]$                                                                                                    \\
                     &     & $[S_{3,xz^2-xy^2}^{T_1^-}, S_{3,yz^2-yx^2}^{T_1^-}, S_{3,zx^2-zy^2}^{T_1^-}]$ & $ \left[ Q^{2\dagger} \begin{pmatrix} 0 & P_{10} \\ P_{10}^* & 0 \end{pmatrix} Q^2, \, Q^\dagger \begin{pmatrix} 0 & P_{10} \\ P_{10}^* & 0 \end{pmatrix} Q, \, \begin{pmatrix} 0 & P_{10} \\ P_{10}^* & 0 \end{pmatrix} \right] $                                                                                   \\
                     &     & $[S_{4,xz^2-xy^2}^{T_1^-}, S_{4,yz^2-yx^2}^{T_1^-}, S_{4,zx^2-zy^2}^{T_1^-}]$ & $ \left[ Q^{2\dagger} \begin{pmatrix} 0 & i P_{10} \\ -i P_{10}^* & 0 \end{pmatrix} Q^2, \, Q^\dagger \begin{pmatrix} 0 & i P_{10} \\ -i P_{10}^* & 0   \end{pmatrix} Q, \, \begin{pmatrix} 0 & i P_{10} \\ -i P_{10}^* & 0 \end{pmatrix} \right]   $                                                                \\
                     &     &                                                                                                                                                                                                                                                                                                                                                                                                      \\

            \hline
                     &     &                                                                                                                                                                                                                                                                                                                                                                                                      \\

            $T_2$    & $+$ & $[S_{1,x}^{T_2^-}, S_{1,y}^{T_2^-}, S_{1,z}^{T_2^-}]$                         & $ \left[ \begin{pmatrix} P_1 & 0 \\ 0 & P_1^* \end{pmatrix}, \,  \begin{pmatrix} P_2 & 0 \\ 0 & P_2^* \end{pmatrix}, \, \begin{pmatrix} P_3 & 0 \\ 0 & P_3^* \end{pmatrix} \right] $                                                                                                                                 \\
                     &     & $[S_{2,x}^{T_2^-}, S_{2,y}^{T_2^-}, S_{2,z}^{T_2^-}]$                         & $ \left[ Q^{2\dagger} \begin{pmatrix} P_7 & 0 \\ 0 & P_7^* \end{pmatrix} Q^2,\, Q^\dagger \begin{pmatrix} P_7 & 0 \\ 0 & P_7^* \end{pmatrix} Q,\, \begin{pmatrix} P_7 & 0 \\ 0 & P_7^* \end{pmatrix} \right] $                                                                                                       \\
                     & $-$ & $[S_{1,x}^{T_2^-}, S_{1,y}^{T_2^-}, S_{1,z}^{T_2^-}]$                         & $ \left[ \begin{pmatrix} P_1 & 0 \\ 0 & -P_1^* \end{pmatrix}, \, \begin{pmatrix} P_2 & 0 \\ 0 & -P_2^* \end{pmatrix}, \, \begin{pmatrix} P_3 & 0      \\0   & -P_3^*\end{pmatrix} \right] $                                                                                                                          \\
                     &     & $[S_{2,x}^{T_2^-}, S_{2,y}^{T_2^-}, S_{2,z}^{T_2^-}]$                         & $ \left[ Q^{2\dagger} \begin{pmatrix}P_7 & 0      \\ 0   & -P_7^*\end{pmatrix} Q^2,\, Q^\dagger \begin{pmatrix}P_7 & 0 \\ 0   & -P_7^*\end{pmatrix} Q,\, \begin{pmatrix}P_7 & 0      \\ 0   & -P_7^*\end{pmatrix} \right]            $                                                                               \\
                     &     & $[S_{3,x}^{T_2^-}, S_{3,y}^{T_2^-}, S_{3,z}^{T_2^-}]$                         & $ \left[ Q^{2\dagger} \begin{pmatrix}0 & P_9 \\P_9^* & 0\end{pmatrix} Q^2,\,Q^\dagger \begin{pmatrix}0 & P_9 \\P_9^* & 0\end{pmatrix} Q,\,\begin{pmatrix}0     & P_9 \\P_9^* & 0\end{pmatrix} \right] $                                                                                                              \\
                     &     & $[S_{4,x}^{T_2^-}, S_{4,y}^{T_2^-}, S_{4,z}^{T_2^-}]$                         & $ \left[ Q^{2\dagger} \begin{pmatrix} 0 & i P_9 \\-i P_9^* & 0\end{pmatrix}  Q^2,\, Q^{\dagger} \begin{pmatrix} 0 & i P_9 \\-i P_9^* & 0\end{pmatrix}  Q, \, \begin{pmatrix} 0 & i P_9 \\-i P_9^* & 0\end{pmatrix} \right]                                                                                   $       \\
                     &     & $[S_{5,x}^{T_2^-}, S_{5,y}^{T_2^-}, S_{5,z}^{T_2^-}]$                         & $ \left[ Q^{2\dagger} \begin{pmatrix} 0 & P_{11} \\P_{11}^* & 0\end{pmatrix}  Q^2,\,Q^\dagger \begin{pmatrix} 0 & P_{11} \\P_{11}^* & 0\end{pmatrix}  Q, \, \begin{pmatrix} 0 & P_{11} \\P_{11}^* & 0\end{pmatrix} \right]                 $                                                                         \\
                     &     & $[S_{6,x}^{T_2^-}, S_{6,y}^{T_2^-}, S_{6,z}^{T_2^-}]$                         & $ \left[ Q^{2\dagger} \begin{pmatrix}0 & i P_{11} \\-iP_{11}^* & 0\end{pmatrix}   Q^2,\,Q^\dagger \begin{pmatrix}0 & i P_{11} \\-iP_{11}^* & 0\end{pmatrix} Q,\, \begin{pmatrix}0 & i P_{11} \\-iP_{11}^* & 0\end{pmatrix} \right]                                                                                 $ \\
            \label{tab_Td_basis_2}
        \end{tabular}
    \end{ruledtabular}
\end{table*}

\bibliographystyle{apsrev4-2}
\bibliography{main.bib}

\clearpage
\end{document}